\documentclass[twocolumn]{aastex62}
\begin{document}
\title{The Absolute Magnitude of the Sun in Several Filters} 

\correspondingauthor{Christopher N. A. Willmer}
\email{cnaw@as.arizona.edu}

\author[0000-0001-9262-9997]{Christopher N. A. Willmer}
\affiliation{Steward Observatory, University of Arizona \\
933 North Cherry Avenue \\
Tucson, AZ 85721, USA}
\received{2018 04 04}
\revised{2018 04 18}
\accepted{2018 04 19}
\submitjournal{\apjs}


\begin{abstract}
This paper presents a table with estimates of the absolute magnitude
of the Sun and the conversions from $vegamag$ to the AB and ST systems
for several wide-band 
filters used in ground and space-based observatories. These estimates
use the dustless spectral energy distribution (SED) of Vega, calibrated
absolutely using the SED of Sirius, to set the $vegamag$ zero-points and a 
composite spectrum of the Sun that coadds space-based observations
from the ultra-violet to the near infrared with models of the Solar
atmosphere. The uncertainty of the absolute magnitudes is
estimated comparing the synthetic colors
with photometric measurements of solar analogs and is found to be
$\sim$ 0.02 magnitudes. Combined with the uncertainty of $\sim$ 2\% in
the calibration of the Vega SED, the errors of these absolute
magnitudes are $\sim$ 3--4\%. Using these SEDs,
for the three of the most utilized filters in extragalactic work 
the estimated absolute magnitudes of the Sun 
are $M_B$ = 5.44, $M_V$ = 4.81 and $M_K$ = 3.27 mag \added{in the $vegamag$
system and $M_B$ = 5.31, $M_V$ = 4.80 and $M_K$ = 5.08 mag in AB}.

\end{abstract}
%
\section{Introduction}

Several astrophysical quantities, such as the masses and
luminosities of stars and galaxies are often described in terms of
solar units. The luminosity density (the integral of the luminosity
function) is even more specific as it is usually expressed in terms of 
solar luminosities within a given photometric band (e.g., $B$ or $K$).
\added{
The consistent absolute calibration of flux measurements
is still an essential endeavor in astrophysics, because of the expansion
of wavelength coverage and ever increasing sensitivity of
instruments both from the ground and space (see
\citet{Bohlinetal2014} for a comprehensive review).
Because the first catalogs of stellar photometry used
Vega as the prime calibrator
\citep{1953ApJ...117..313J, 1955AnAp...18..292J, Johnson1966araa}, 
magnitudes are commonly referred to that star.
However, to overcome the effects of dust and molecular lines on
stellar spectra which are difficult to model, there has been a shift
to adopt either the AB system of \citet{Oke_Gunn_1983}, where the
calibrating spectrum is flat in $f_\nu$ or the ST system
\citep{Bessell1998, synphot}, for a flat spectrum in $f_\lambda$. 
Both, in their turn, can be referred to
observations of white dwarfs, which are calibrated through stellar
models and ultimately through the use of laboratory reference standards
\citep{Bohlinetal2014}. 
}

Previous compilations of the Sun's absolute magnitude were
published by \citet{GalacticAstronomy} for the Johnson-Cousins-Glass system
and \citet{Blanton2003} for the Sloan Digital Sky Survey (SDSS)
filters redshifted to z=0.1 in AB magnitudes. \citet{Engelke2010}
calculate the $apparent$ magnitude of the Sun for
several filters including the Johnson-Cousins ($UBVRI$), 2MASS
($JHK$), and Spitzer IRAC 8$\mu$m and MIPS 24 $\mu$m which can be
easily converted into absolute magnitudes.
The conversion constants between the
$vegamag$ system, where the absolute calibration is referred to Vega
and the AB \citep{Oke_Gunn_1983} and ST \citep{Bessell1998, synphot}
systems for different filters are less common to find, and
the most extensive compilation of the $vegamag$ to AB measurements
was published by \citet{Fukugita1995}.
The aim of
this paper is to provide a handy reference for the absolute magnitude
of the Sun in several filters used primarily by large surveys, and the 
additive constants (i.e., the magnitude of Vega) that transform
$vegamag$ into the  AB and ST systems. This is done using recent
determinations of the spectral energy distribution (SED) of Vega and
the Sun derived from space-based  observations combined with models of
the atmospheres of these stars.

This paper is organized as follows: Section 2 describes the
filter curves, the measurement of synthetic magnitudes and the
determination of the $vegamag$ zero-points; Section 3
describes the construction of the solar spectrum and a summary
concludes in Section 4.

\section{Filter Curves and  Synthetic magnitudes}

The filter profiles were compiled from the literature,
e.g., \citet{Tonry2012, 2015PASP..127..102M}, or downloaded
from the databases of observatories or surveys, e.g., JWST, Dark Energy
Survey. The filter profiles include the throughput due
to the telescope, instrument optics and detector quantum
efficiency (e.g., HST and JWST filters). In the case of HST filters,
the latest files available in the
$synphot$
database \footnote{\url{http://www.stsci.edu/hst/observatory/crds/throughput.html}}
were used.
Most of the filters used in
the ground-based observations, e.g., Sloan Digital Sky Survey
\citep{SDSS},  Pan-STARRS \citep{Tonry2012},
also include a contribution due to the Earth's atmosphere. 
As the JWST Mid Infra Red Instrument (MIRI) filter response curves of
\citet{GlasseMIRI} only contain the instrument throughput, these
were multiplied by the  expected JWST mirror reflectance as provided
by the STScI NIRCam Team.

The \hfill reconstruction of the full system throughput
\hfill using CCD photometry for the  $U$, $B$, $V$ bands~ of\\
\citet{1953ApJ...117..313J} and
\citet{1955AnAp...18..292J}, and $R$ and $I$ of \citet{Cousins1976},
which were measured using photo-electric photometers
has been addressed in several works among which
\citet{2006AJ....131.1184M}, \citet[hereafter BM12]{BM12} and
\citet{2015PASP..127..102M}.
In the latter work, the authors re-determine the profiles of 39
filters ($U$, $B$, $V$, $R$ and $I$ among them) 
using spectroscopic libraries from the HST/STIS and IRTF/SPEX
instruments, which provide coverage from the UV \replaced{bluewards}{shortwards} of the
atmospheric cutoff to the Near-Infrared. In their comparison with BM12,
\citet{2015PASP..127..102M} 
find agreement within 2\% for most filters, a notable exception
being $U$, which shows a 5\% difference, which they trace to the
use by BM12 of the MILES library \citep{Falcon2011},
which has a less extensive $U$ coverage than the  STIS spectroscopy
used by \citet{2015PASP..127..102M}. 

The wavelength limits of filter curves adopted in this work are set by
the wavelengths where the system throughput reaches below 10$^{-4}$ of the
peak value. The filters are normalized by the maximum value and then
resampled using linear interpolation as using spline
interpolations can introduce spurious features in
filters that do not have smooth curves (e.g., 2MASS).


The calculation of synthetic magnitudes follows BM12 eq. A11:
\begin{equation}
magnitude = -2.5 log_{10} \bigg[ \frac{\int f_\lambda(\lambda) R(\lambda)
 \lambda d\lambda}{\int R(\lambda) \lambda d\lambda} \bigg] - zp
\end{equation}
where $f_\lambda(\lambda)$ is the stellar flux density in
erg~cm$^{-2}$~s$^{-1}$~\AA$^{-1}$, $R(\lambda)$ the product of the  
detector quantum efficiency $\times$ filter throughput $\times$ 
unitless fractional transmission of the
total telescope optical train and $zp$ is the zero-point correction for a
given magnitude system. The integral is calculated at each filter
wavelength by determining the stellar flux value using linear interpolation.

The AB system is defined such that the zero-point flux density for every
filter is 3631 Jy, corresponding to a  $zp$ = 48.60. For the ST
system the  $zp$ = 21.10 and is defined such that the magnitude of
Vega in the (Johnson) V band is +0.03 \citep{Bessell1998}.
In both cases these zero-points assume the standard calibration
spectrum is flat either in frequency (AB) or wavelength (ST) \citep{synphot}.

In the case of the \citet{Johnson1966araa} $UBVRI$ or $vegamag$
\citep{synphot} system, the zero point is defined from the colors of
several A stars, and because of this, Vega has a small magnitude offset in
all bands that must be accounted for when using
its spectrum as a flux standard \citep{Rieke2008}.
However, the finding that Vega's spectrum shows the
presence of a debris disk 
(\citet{Aumann1984, Rieke2008, Su2013, 2014AJ....147..127B}), 
and that in addition is a rapid rotator \citep{Peterson2006},
limits the ability of theoretical models of matching its SED,
and have prompted the search of other AV stars to serve as spectral flux
standards, e.g., \citet{Cohen1992, Bessell1998, Engelke2010,
  2014AJ....147..127B}.

\begin{deluxetable*}{crrDDrr}[t]
\tabletypesize {\scriptsize}
\tablecaption{Template Colors\label{tab:Astars}}
\tablehead{
\colhead{Color}     &
\colhead{Sirius\tablenotemark{a}} & 
\colhead{\citet{Bessell1998} }  &
\multicolumn2c{BD+60~1753\tablenotemark{b}} & 
\multicolumn2c{IRSA\tablenotemark{c}}   &
\colhead{\citet{Rieke2008} }  &
\colhead{\citet{Engelke2010}} \\
(1) & (2)~~~~~ & (3)~~~~~~~~~~~ & {}& (4) & {}& (5) & (6)~~~~~~~~~ & (7)~~~~~~~~
}
\decimals
\startdata
U   - B               &-0.054 &-0.045~~~~~~~~~~& 0.008  & {}             &  0.022~~~~~~&-0.029~~~~~~\\
B   - V               &-0.015 &-0.01~~~~~~~~~~~& 0.023\tablenotemark{d} & 0.080\tablenotemark{d,e} &  0.001~~~~~~&-0.005~~~~~~\\
V   - R               &-0.013 &-0.012~~~~~~~~~~& 0.007  & {}             & -0.006~~~~~~& 0.010~~~~~~\\
R   - I               &-0.016 &-0.008~~~~~~~~~~& 0.009  & {}             & -0.005~~~~~~& 0.011~~~~~~\\
V   - 2MASS\_K        &-0.089 &-0.061\tablenotemark{f}~~~~~~~~~& 0.028  & {}             & -0.028~~~~~~&-0.025~~~~~~\\
2MASS\_J  - 2MASS\_H  &-0.015 &-0.018\tablenotemark{f}~~~~~~~~~& 0.003  &-0.039~~~~      & -0.004~~~~~~& 0.008~~~~~~\\
2MASS\_H  - 2MASS\_Ks &-0.006 &-0.009\tablenotemark{f}~~~~~~~~~& 0.002  & 0.006~~~~      & -0.003~~~~~~&-0.019~~~~~~\\
IRAC\_3.6 - IRAC\_4.5 &-0.002 & {}             &-0.000~~~  & 0.013      &  0.001~~~~~~&-0.003~~~~~~\\
IRAC\_4.5 - IRAC\_5.8 &-0.002 & {}             &-0.000~~~  &-0.006     & -0.000~~~~~~&-0.005~~~~~~\\
IRAC\_5.8 - IRAC\_8.0 &-0.003 & {}             &-0.002~~~  &-0.010     & -0.001~~~~~~&-0.001~~~~~~\\
WISE\_1   - WISE\_2   &-0.003 & {}             &-0.000~~~  &-0.027     &  0.000~~~~~~&-0.005~~~~~~\\
WISE\_2   - WISE\_3   &-0.008 & {}             &-0.005~~~  &-0.025     & -0.001~~~~~~&-0.002~~~~~~\\
WISE\_3   - WISE\_4   &-0.006 & {}             &-0.004~~~  & 0.596\tablenotemark{g} &  0.000~~~~~~& 0.010~~~~~~\\
\enddata
\tablenotetext{a}{\it{sirius\_stis\_002.fits}} 
\tablenotetext{b}{\citet{Spitzer} archive}
\tablenotetext{c}{\it{bd60d1753\_stis\_004.fits}}
\tablenotetext{d}{Tycho filters \citep{Hog2000}}
\tablenotetext{e}{\citet{Hog2000}}
\tablenotetext{f}{\citet{Carter1990} SAAO system}
\tablenotetext{g}{low S/N measurement in WISE\_4}
\end{deluxetable*}

The use of Sirius as a flux standard for the infrared was initially proposed
by \citet{Cohen1992}, and adopted by \citet{Bessell1998} and
\citet{Engelke2010}. A detailed analysis of the SED of Sirius was done by
\citet{2014AJ....147..127B} who created a template combining IUE and HST/STIS
spectra for wavelengths between $\sim$ 0.15~$\mu$m and
1.0~$\mu$m with a Kurucz model of Sirius to 300~$\mu$m. 
\citet{2014AJ....147..127B} found the STIS measurements agree to
better than 1\% with the Kurucz model and that this model also
shows good agreement ($\sim$ 2\%) with infrared photometry obtained by the
Midcourse Space Experiment (MSX) satellite.
Based on these results, \citet{2014AJ....147..127B} concluded that
Sirius can be used as standard calibrator for the infrared
and its composite spectrum is available in the
$CALSPEC$ database ({\it{sirius\_stis\_002.fits}}).
Once adopting Sirius as the flux standard, \citet{2014AJ....147..127B}
re-normalized the Vega composite dust-free template spectrum that
combines IUE and STIS observations of Vega with two Kurucz models for
Vega with T=9550K (for the extreme UV) and T=9400K (for the
visible-far IR) \citep{2014AJ....147..127B}, which is file
{\it{alpha\_lyr\_stis\_008.fits}} in the $CALSPEC$ database.

\begin{figure*}[t]
\includegraphics{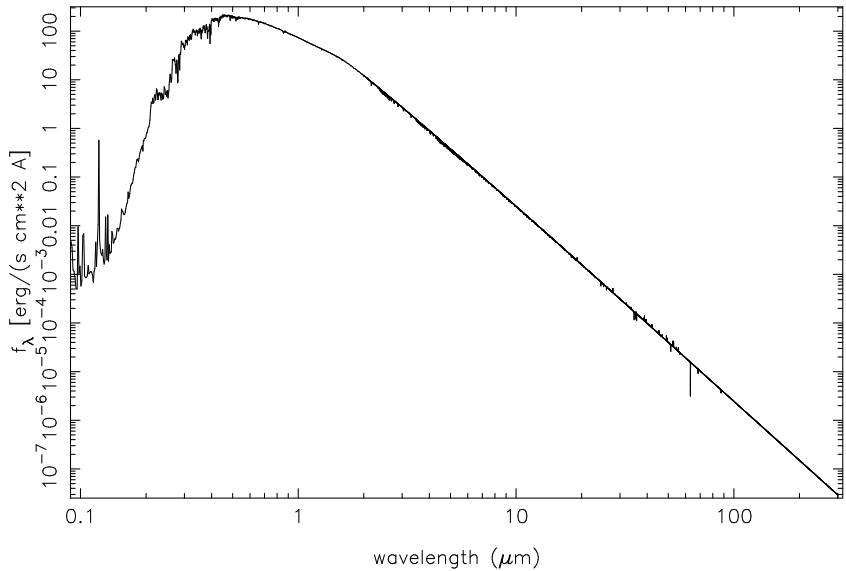}
\vspace{7.5cm}
\caption{Composite spectrum of the Sun, combining the observed
  spectrum of \citet{2017JGRA..122.5910H} to 2.0~$\mu$m, the
  \citet{Fontenla2011} 
  model between 2.0~$\mu$m and 100~$\mu$m and the Kurucz model
  {\it{sun\_mod\_001.fits}} from 100~$\mu$m to 300~$\mu$m. The data
  used to create this figure are available in the online journal.}
\label{fig:sol}
\end{figure*}

Other AV templates have been defined using models and
observations. \citet{Rieke2008} constructed a dustless A0V template
using the \citet{Kurucz2005} model of Vega and normalizing the
spectrum in the infrared after correcting for the contribution of
the debris disk. By means of a detailed comparison with the photometry
of A dwarfs and solar analogs \citet{Rieke2008} showed that this A0V
template as well as the solar SED they calculated in the same paper
give consistent calibrations the infrared.
An AV template combining ground-based observations of 109 Vir with the average
NICMOS observations for eight A type stars, ISO observations of Sirius
(from 2.4 to 9.4 $\mu$m) and beyond 9.4 $\mu$m, a Kurucz model spectrum
for Sirius was compiled by \citet{Engelke2010} who find that the
calibration uncertainties are $\lesssim$ 2\%.
The final template considered here is the A1V star BD+60~1753, which
is one of the $IRAC$ calibrators \citep{Reach2005} that has a
$CALSPEC$ spectrum {\it{bd60d1753\_stis\_004.fits}} which combines HST/STIS
observations from  1140~\AA~ to 10120~\AA~ with \replaced{BOZS}{BOSZ} 
models beyond
10120~\AA~ \citep{Bohlin2017}.

A comparison between colors measured using the
\citet{2014AJ....147..127B} {\it{alpha\_lyr\_stis\_008.fits}} 
spectrum of Vega as standard (which will be zero by
definition) with those of  the AV templates discussed above is shown Table
\ref{tab:Astars}. Column (1) identifies the photometric color,
column(2) the  synthetic color measured using the Sirius spectrum of
\citet{2014AJ....147..127B} 
followed in column (3) by photometric measurements \added{of Sirius} in \citet{Bessell1998}. 
Column (4) shows the synthetic photometry colors for BD+60~1753, while
column (5) shows measurements for this star available in \citet{Hog2000} 
and the \citet{Spitzer} archive. Columns (6) and (7) show the
synthetic \replaced{colores}{colors measured for the}
\citet{Rieke2008} and \citet{Engelke2010} templates respectively.
The average difference
between the synthetic and observed colors of Sirius and BD+60~1753 
are -0.006 $\pm$ 0.010 and and 0.007 $\pm$ 0.028 respectively.
The mean difference between the synthetic colors measured for the four
templates and the Vega SED are $\lesssim$ 0.018 magnitudes and with
a dispersion of the same order of magnitude ($\lesssim$ 0.024
magnitudes). These results suggest that the calibration uncertainty
introduced by using the $CALSPEC$ spectrum of Vega is $\sim$ 2\%.

In this work the $vegamag$ magnitudes are calculated using the Vega SED
of \citet{2014AJ....147..127B} ({\it{alpha\_lyr\_stis\_008.fits}} in
the STScI $CALSPEC$ database), assuming a Vega magnitude of V=0.03 (BM12).

\section{The Solar Spectrum}

The solar SED used here also combines observations with model spectra.
The observed spectrum is a composite calculated by \citet{2017JGRA..122.5910H}
using data from over 20 space-based instruments 
for an arbitrary date (2008-Dec-19, JDN=2454820) during the solar
minimum\replaced{, though there are minimal changes in the spectrum for other
dates at the same epoch.} {. Spectra for other dates around the solar
  minimum show no significant change relative to the spectrum adopted here.}
\citet{2017JGRA..122.5910H} use a probabilistic approach to 
combine observations at each time step weighting the spectra
by their uncertainties and accounting for fluctuations over
time between different instruments at the same wavelength. The absolute
calibration is set by using the $ATLAS~3$ composite spectrum of
\citet{Thuillier2004}, and constraining the Total Solar
Irradiance (TSI) to the value measured for each day by
\citet{2017GeoRL..44.1196D}. 
The observed composite ends at $\sim$ 2.0~$\mu$m, and to extend the SED
into the infrared the model spectra of \citet{Fontenla2011}
and \citet{Kurucz2011}
are used.
The \citet{Fontenla2011} model 
uses the Solar Irradiance Physical Modeling (SRPM) system to produce
solar irradiance spectra from 0.012~$\mu$m to 100~$\mu$m through a
combination of non-LTE models with semi-empirical physical models
derived from observed spectra to produce the solar SED.
The \citet{Fontenla2011} spectrum was scaled by
the bi-weight average \citep{Beers1990} ratio between the
\citet{2017JGRA..122.5910H} composite and the model for wavelengths
between 1.8~$\mu$m and 2.0~$\mu$m (1.0299 $\pm$0.0074). Beyond 100~$\mu$m
the special Kurucz model at R=5000 calculated for the $CALSPEC$
database ({\it{sun\_mod\_001.fits}}, \citet{Kurucz2011})  is used, and
to eliminate any 
\replaced{steps}{discontinuities} in the transition between source
spectra, the \added{bi-weight} ratio between  
the re-normalized \citet{Fontenla2011} and Kurucz models calculated
between 90~$\mu$m and 100.0~$\mu$m (0.9940 $\pm$ 0.0073), was used to scale the latter.
Fig~\ref{fig:sol} shows the composite spectrum.

\begin{figure*}
\includegraphics{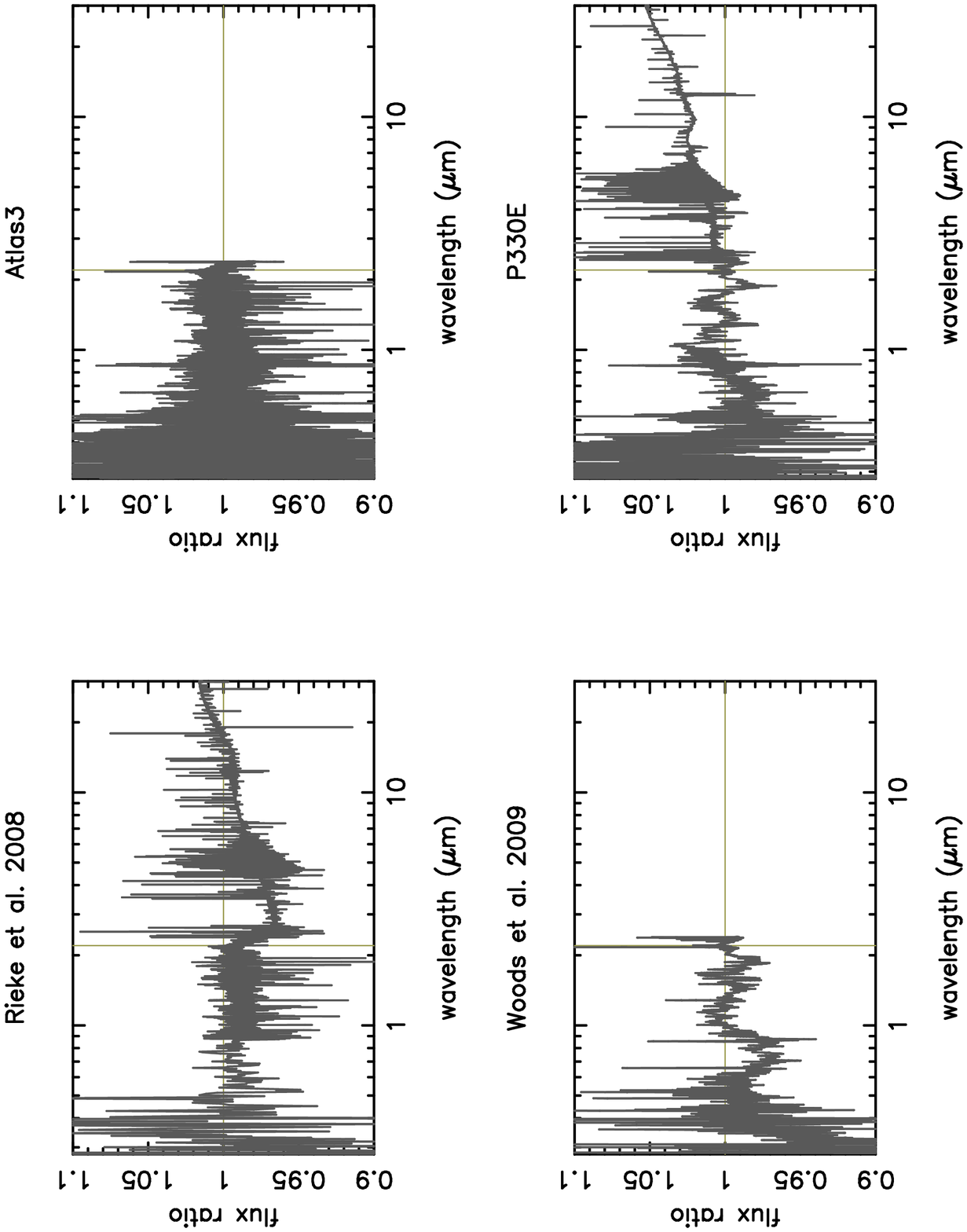}
\vspace{11.5cm}
\caption{Ratio between published SEDs of the Sun and the composite
  used here. The panels show the ratio for the \citet{Rieke2008},
  ATLAS-3 \citep{Thuillier2004}, \citet{Woods2009} and the solar
  analog P330E from the \citet{calspec} database, normalized to have
  the same flux as the composite spectrum at 2.2~$\mu$m. The vertical line is
  located at $\lambda$ = 2.2~$\mu$m. \added{The spikes seen in these
    ratios are caused by small mismatches in the wavelengths and
    resolutions of the spectra. Differences of the order of $\sim$ 5\%
    can be seen in the rations between spectra. The \citet{Rieke2008}
    spectrum between 2.2 $\mu$m and 20 $\mu$m where the \citet{Engelke1992}
    approximation is used is systematically fainter than the
    \cite{Fontenla2011} models, while shortward of 2.2 $\mu$m the
    agreement with the composite adopted here is very good. The
    spectrum of P330E is systematically brighter than the composite
    used here for wavelengths longer than $\sim$ 2 $\mu$m.}
}
\label{fig:sol_comp}
\end{figure*}

Figure~\ref{fig:sol_comp} shows the ratios between this composite
spectrum with other determinations in the literature --
\citet{Rieke2008, Thuillier2004, Woods2009}.
Also shown is a comparison with the solar analog P330E after scaling its
spectrum to have the same flux density as the composite at 2.2~$\mu$m
(8.58593 erg cm$^{-2}$sec$^{-1}$\AA$^{-1}$).
The ratios being plotted are calculated at each tabulated wavelength
of the source spectrum and  
estimating the composite's flux using linear interpolation.
The SED ratios are very close to 1 (0.998, 1.005, 0.993 and 1.017 for
\citet{Rieke2008,Thuillier2004,Woods2009} and P330E respectively) and
dispersions of 0.037 or better, in all cases using the bi-weight estimator.
The rms fluctuations range from 0.02 for the \citet{Rieke2008} solar
spectrum, to 0.22 in the case of P330E.

\twocolumngrid
The comparison between colors estimated using the solar composite
spectrum with measurements by\\  
\citet{Ramirez2012} and
\citet{Casagrande2012} of solar analogs is presented in Table
\ref{tab:solar_colors_vega}. The average difference in colors for
(composite - solar analogs) is  -0.018 $\pm$ 0.030 magnitudes, suggesting
that the composite spectrum shows consistent measurements both in the
UV--visible and the infrared.
Table \ref{tab:solar_colors_vega} also compares the colors of the new
composite with
measurements using the \citet{Rieke2008} solar model which combines
the \citet{Thuillier2003} SED with the \citet{Engelke1992}
approximation for the near to mid infrared. The (composite - \citet{Rieke2008})
differences are  -0.003 $\pm$ 0.013 magnitudes, though 
\deleted{it should be noted that}
these measurements may not be completely independent as
\citet{2017JGRA..122.5910H} use the \citet{Thuillier2004} spectrum to
set the absolute calibration of the solar spectrum.

\begin{deluxetable*}{crrrrrr}
\tabletypesize {\scriptsize}
\tablecaption{Colors of the solar composite compared to solar analogs\label{tab:solar_colors_vega}}
\tablehead{
\colhead{Color}     &
\colhead{Composite} & 
\colhead{Solar Analogs} &
\colhead{difference} & 
\colhead{\citet{Rieke2008} SED} & 
\colhead{difference}\\
}
\decimals
\startdata
U - B         & 0.164 & 0.166\tablenotemark{a}~~~~~~ & -0.002 & 0.138~~~~~~&  0.026\\
B - V         & 0.629 & 0.653\tablenotemark{a}~~~~~~ & -0.024 & 0.629~~~~~~&  0.000\\
V - R         & 0.387 & 0.352\tablenotemark{a}~~~~~~ &  0.035 & 0.388~~~~~~& -0.001\\
V - I         & 0.712 & 0.702\tablenotemark{a}~~~~~~ &  0.010 & 0.717~~~~~~& -0.005\\
V - 2MASS\_J  & 1.145 & 1.198\tablenotemark{b}~~~~~~ & -0.053 & 1.143~~~~~~& 0.002\\
V - 2MASS\_H  & 1.494 & 1.484\tablenotemark{b}~~~~~~ &  0.010 & 1.492~~~~~~& 0.002\\
V - 2MASS\_Ks & 1.542 & 1.560\tablenotemark{b}~~~~~~ & -0.018 & 1.545~~~~~~& -0.003\\
V - WISE\_1   & 1.553 & 1.608\tablenotemark{b}~~~~~~ & -0.055 & 1.530~~~~~~& 0.023\\
V - WISE\_2   & 1.530 & 1.563\tablenotemark{b}~~~~~~ & -0.033 & 1.515~~~~~~& 0.015\\
V - WISE\_3   & 1.549 & 1.552\tablenotemark{b}~~~~~~ & -0.003 & 1.551~~~~~~& -0.002\\
V - WISE\_4   & 1.539 & 1.604\tablenotemark{b}~~~~~~ & -0.065 & 1.559~~~~~~& -0.020\\
\enddata                                             
\tablenotetext{a}{\citet{Ramirez2012}}
\tablenotetext{b}{\citet{Casagrande2012}}                                     
\end{deluxetable*}

The estimated errors in the solar magnitudes change as a function of
wavelength due to the uncertainty on the absolute calibration using
the Vega and Solar SED  $\sim$~2--3\% \citep{Rieke2008,
  2014AJ....147..127B}. These added in quadrature result in
uncertainties $\sim$ 3\% over the range covered by solar analogs.
These can become larger ($\sim$ 5\%) as one transitions towards the
Mid-Infrared due to the difficulty in calibrating the space-based
instruments in this wavelength range, e.g., \citet{Fontenla2011}.

To derive the Sun's absolute magnitudes, the
IAU 2012 definitions of the astronomical unit (AU) \citep{Prsa2016}
and parsec
were used, giving a distance modulus for the Sun of -31.5721 magnitudes.
To rationalize the use of solar constants, the IAU
in 2015 adopted a nominal value for the Sun's luminosity L$_{odot}$ =
3.828$\times$10$^8$ W \citep{Prsa2016}, which corresponds to an
average TSI of 1361 W m$^{-2}$ at 1 AU and an absolute bolometric
magnitude of $M_{Bol}$=4.74. 

Table~\ref{tab:1} lists in columns (1) the filter,
(2), (3) and (4) the absolute magnitude of the Sun in the $vegamag$,
AB and ST systems, in columns (5) and (6) and (7) the apparent
magnitude in $vegamag$, AB, ST; in columns (8) and (9) tabulates the offsets
between the $vegamag$ and AB  and $vegamag$ and ST systems
followed in column (10) by the pivot wavelength 
and in column (11) the source of the throughput curves, identified in
the table notes.  

\startlongtable
\begin{deluxetable*}{lrrrrrrrrrc}
\tablecaption{Magnitudes of the Sun\label{tab:1}}
\tablehead{
\colhead{Filter} & \colhead{Abs} &\colhead{Abs} &\colhead{Abs} &
 \colhead{App} & \colhead{App} & \colhead{App} &
\colhead{Vega} & \colhead{Vega} &
\colhead{$\lambda_{pivot}$} &\colhead{Source} \\
{} & (Vega) & (AB) & (ST) &  (Vega) & (AB) & (ST) & (AB) &
(ST)&$\mu$m~ & {}\\
~~~~~~~~~(1) & (2)~ &(3)~ &(4)~ &(5)~ &(6)~ &(7)~ &(8)~ &(9)~ &(10) &(11)
}
\startdata
Johnson\_U &    5.61 &    6.33 &    5.42 &  -25.97 &  -25.25 &  -26.15 &    0.721 &   -0.183 & 0.3611 & 1\\
Johnson\_B &    5.44 &    5.31 &    4.84 &  -26.13 &  -26.26 &  -26.74 &   -0.128 &   -0.605 & 0.4396 & 1\\
Johnson\_V &    4.81 &    4.80 &    4.81 &  -26.76 &  -26.77 &  -26.76 &   -0.013 &    0.001 & 0.5511 & 1\\
Cousins\_R &    4.43 &    4.60 &    5.00 &  -27.15 &  -26.97 &  -26.57 &    0.178 &    0.578 & 0.6582 & 1\\
Cousins\_I &    4.10 &    4.51 &    5.35 &  -27.47 &  -27.06 &  -26.22 &    0.414 &    1.247 & 0.8034 & 1\\
Tycho\_Bt &    5.58 &    5.48 &    4.91 &  -25.99 &  -26.09 &  -26.66 &   -0.097 &   -0.667 & 0.4212 & 1\\
Tycho\_Vt &    4.88 &    4.85 &    4.79 &  -26.69 &  -26.72 &  -26.78 &   -0.035 &   -0.091 & 0.5335 & 1\\
Hipparcos\_Hp &    4.87 &    4.87 &    4.88 &  -26.70 &  -26.70 &  -26.69 &   -0.002 &    0.011 & 0.5508 & 1\\
2MASS\_J &    3.67 &    4.54 &    6.31 &  -27.90 &  -27.03 &  -25.26 &    0.870 &    2.644 & 1.2393 & 2\\
2MASS\_H &    3.32 &    4.66 &    7.06 &  -28.25 &  -26.91 &  -24.51 &    1.344 &    3.739 & 1.6495 & 2\\
2MASS\_Ks &    3.27 &    5.08 &    8.07 &  -28.30 &  -26.49 &  -23.50 &    1.814 &    4.798 & 2.1638 & 2\\
SDSS\_u &    5.49 &    6.39 &    5.45 &  -26.08 &  -25.18 &  -26.12 &    0.900 &   -0.037 & 0.3556 & 3\\
SDSS\_g &    5.23 &    5.11 &    4.78 &  -26.34 &  -26.47 &  -26.80 &   -0.125 &   -0.456 & 0.4702 & 3\\
SDSS\_r &    4.53 &    4.65 &    4.91 &  -27.04 &  -26.93 &  -26.66 &    0.119 &    0.380 & 0.6176 & 3\\
SDSS\_i &    4.19 &    4.53 &    5.21 &  -27.38 &  -27.05 &  -26.37 &    0.332 &    1.012 & 0.7490 & 3\\
SDSS\_z &    4.01 &    4.50 &    5.57 &  -27.56 &  -27.07 &  -26.00 &    0.494 &    1.560 & 0.8947 & 3\\
DES\_u &    5.83 &    6.14 &    5.38 &  -25.74 &  -25.44 &  -26.20 &    0.307 &   -0.452 & 0.3859 & 4\\
DES\_g &    5.17 &    5.05 &    4.78 &  -26.41 &  -26.52 &  -26.80 &   -0.114 &   -0.391 & 0.4820 & 4\\
DES\_r &    4.45 &    4.61 &    4.96 &  -27.12 &  -26.96 &  -26.61 &    0.159 &    0.505 & 0.6423 & 4\\
DES\_i &    4.14 &    4.52 &    5.29 &  -27.43 &  -27.05 &  -26.28 &    0.382 &    1.152 & 0.7807 & 4\\
DES\_z &    4.01 &    4.50 &    5.62 &  -27.56 &  -27.07 &  -25.95 &    0.493 &    1.610 & 0.9158 & 4\\
DES\_Y &    3.96 &    4.50 &    5.78 &  -27.61 &  -27.07 &  -25.79 &    0.540 &    1.819 & 0.9866 & 4\\
PS1\_g &    5.14 &    5.03 &    4.77 &  -26.43 &  -26.54 &  -26.80 &   -0.112 &   -0.376 & 0.4849 & 5\\
PS1\_r &    4.53 &    4.64 &    4.92 &  -27.05 &  -26.93 &  -26.66 &    0.120 &    0.390 & 0.6201 & 5\\
PS1\_i &    4.18 &    4.52 &    5.22 &  -27.39 &  -27.05 &  -26.35 &    0.339 &    1.033 & 0.7535 & 5\\
PS1\_z &    4.02 &    4.51 &    5.50 &  -27.55 &  -27.07 &  -26.07 &    0.483 &    1.482 & 0.8674 & 5\\
PS1\_Y &    3.99 &    4.50 &    5.73 &  -27.59 &  -27.07 &  -25.85 &    0.515 &    1.741 & 0.9628 & 5\\
cfhtls\_u &    5.70 &    6.04 &    5.25 &  -25.87 &  -25.53 &  -26.33 &    0.336 &   -0.455 & 0.3803 & 6\\
cfhtls\_g &    5.15 &    5.03 &    4.77 &  -26.42 &  -26.54 &  -26.80 &   -0.116 &   -0.382 & 0.4844 & 6\\
cfhtls\_r &    4.50 &    4.64 &    4.92 &  -27.07 &  -26.94 &  -26.65 &    0.131 &    0.417 & 0.6248 & 6\\
cfhtls\_i &    4.16 &    4.52 &    5.26 &  -27.41 &  -27.05 &  -26.32 &    0.362 &    1.096 & 0.7678 & 6\\
cfhtls\_z &    4.02 &    4.51 &    5.55 &  -27.56 &  -27.07 &  -26.02 &    0.490 &    1.535 & 0.8859 & 6\\
CFHT\_12kx8k\_B &    5.43 &    5.28 &    4.80 &  -26.14 &  -26.30 &  -26.77 &   -0.157 &   -0.632 & 0.4399 & 7\\
CFHT\_12kx8k\_R &    4.39 &    4.59 &    5.00 &  -27.18 &  -26.98 &  -26.57 &    0.196 &    0.605 & 0.6610 & 7\\
CFHT\_12kx8k\_I &    4.10 &    4.51 &    5.38 &  -27.47 &  -27.06 &  -26.19 &    0.415 &    1.282 & 0.8159 & 7\\
UKIRT\_z &    4.02 &    4.51 &    5.54 &  -27.56 &  -27.07 &  -26.03 &    0.489 &    1.526 & 0.8826 & 8\\
UKIRT\_Y &    3.92 &    4.51 &    5.88 &  -27.66 &  -27.07 &  -25.69 &    0.591 &    1.966 & 1.0315 & 8\\
UKIRT\_J &    3.65 &    4.54 &    6.33 &  -27.92 &  -27.03 &  -25.24 &    0.891 &    2.684 & 1.2502 & 8\\
UKIRT\_H &    3.33 &    4.66 &    7.03 &  -28.25 &  -26.92 &  -24.54 &    1.329 &    3.705 & 1.6360 & 8\\
UKIRT\_K &    3.27 &    5.12 &    8.14 &  -28.30 &  -26.45 &  -23.43 &    1.848 &    4.874 & 2.2060 & 8\\
LSST\_u &    5.65 &    6.27 &    5.40 &  -25.93 &  -25.30 &  -26.17 &    0.627 &   -0.244 & 0.3665 & 9\\
LSST\_g &    5.17 &    5.06 &    4.77 &  -26.40 &  -26.52 &  -26.80 &   -0.116 &   -0.399 & 0.4808 & 9\\
LSST\_r &    4.52 &    4.64 &    4.92 &  -27.05 &  -26.93 &  -26.66 &    0.121 &    0.395 & 0.6210 & 9\\
LSST\_i &    4.18 &    4.52 &    5.22 &  -27.39 &  -27.05 &  -26.35 &    0.340 &    1.034 & 0.7537 & 9\\
LSST\_z &    4.02 &    4.51 &    5.51 &  -27.55 &  -27.07 &  -26.06 &    0.484 &    1.486 & 0.8686 & 9\\
LSST\_y &    3.98 &    4.50 &    5.74 &  -27.59 &  -27.07 &  -25.83 &    0.520 &    1.763 & 0.9705 & 9\\
Bessell\_Murphy\_U &    5.57 &    6.34 &    5.43 &  -26.00 &  -25.23 &  -26.14 &    0.768 &   -0.144 & 0.3597 & 10\\
Bessell\_Murphy\_B &    5.46 &    5.33 &    4.84 &  -26.11 &  -26.24 &  -26.73 &   -0.134 &   -0.620 & 0.4378 & 10\\
Bessell\_Murphy\_V &    4.82 &    4.81 &    4.81 &  -26.75 &  -26.77 &  -26.76 &   -0.017 &   -0.012 & 0.5489 & 10\\
Bessell\_Murphy\_R &    4.44 &    4.61 &    4.99 &  -27.13 &  -26.96 &  -26.58 &    0.168 &    0.548 & 0.6524 & 10\\
Bessell\_Murphy\_I &    4.11 &    4.52 &    5.33 &  -27.46 &  -27.06 &  -26.24 &    0.408 &    1.227 & 0.7984 & 10\\
Bessell\_Murphy\_Bt &    5.60 &    5.51 &    4.93 &  -25.98 &  -26.06 &  -26.65 &   -0.088 &   -0.669 & 0.4190 & 10\\
Bessell\_Murphy\_Vt &    4.89 &    4.86 &    4.79 &  -26.68 &  -26.71 &  -26.79 &   -0.038 &   -0.108 & 0.5300 & 10\\
Bessell\_Murphy\_Hp &    4.93 &    4.92 &    4.86 &  -26.64 &  -26.66 &  -26.71 &   -0.018 &   -0.068 & 0.5349 & 10\\
Bessell\_88\_J &    3.67 &    4.54 &    6.30 &  -27.90 &  -27.03 &  -25.27 &    0.866 &    2.632 & 1.2347 & 11\\
Bessell\_88\_H &    3.32 &    4.66 &    7.05 &  -28.25 &  -26.91 &  -24.52 &    1.337 &    3.726 & 1.6450 & 11\\
Bessell\_88\_K &    3.27 &    5.09 &    8.07 &  -28.30 &  -26.49 &  -23.50 &    1.815 &    4.802 & 2.1663 & 11\\
Bessell\_88\_L &    3.26 &    5.98 &   10.00 &  -28.31 &  -25.59 &  -21.58 &    2.721 &    6.737 & 3.4797 & 11\\
Bessell\_88\_Lprime &    3.26 &    6.17 &   10.39 &  -28.31 &  -25.40 &  -21.18 &    2.914 &    7.135 & 3.8247 & 11\\
Bessell\_88\_M &    3.29 &    6.64 &   11.33 &  -28.28 &  -24.93 &  -20.25 &    3.349 &    8.034 & 4.7347 & 11\\
GALEX\_FUV &   15.22 &   17.30 &   14.54 &  -16.36 &  -14.27 &  -17.03 &    2.085 &   -0.676 & 0.1535 & 12\\
GALEX\_NUV &    8.53 &   10.16 &    8.28 &  -23.04 &  -21.41 &  -23.30 &    1.629 &   -0.253 & 0.2301 & 12\\
WISE\_1 &    3.26 &    5.91 &    9.87 &  -28.31 &  -25.66 &  -21.70 &    2.655 &    6.614 & 3.3897 & 13\\
WISE\_2 &    3.28 &    6.57 &   11.22 &  -28.29 &  -25.00 &  -20.36 &    3.291 &    7.932 & 4.6406 & 13\\
WISE\_3 &    3.26 &    8.48 &   15.28 &  -28.31 &  -23.09 &  -16.29 &    5.215 &   12.019 & 12.5705 & 13\\
WISE\_4 &    3.27 &    9.88 &   17.93 &  -28.30 &  -21.70 &  -13.65 &    6.602 &   14.652 & 22.3142 & 13\\
IRAS12 &    3.26 &    8.30 &   14.89 &  -28.31 &  -23.27 &  -16.69 &    5.037 &   11.621 & 11.3562 & 13\\
IRAS25 &    3.27 &    9.92 &   18.09 &  -28.30 &  -21.65 &  -13.48 &    6.646 &   14.819 & 23.6079 & 13\\
IRAS60 &    3.28 &   11.90 &   22.12 &  -28.29 &  -19.67 &   -9.46 &    8.621 &   18.833 & 60.3699 & 13\\
IRAS100 &    3.29 &   13.14 &   24.47 &  -28.28 &  -18.43 &   -7.10 &    9.854 &   21.186 & 101.1267 & 13\\
IRAC\_3.6 &    3.26 &    6.02 &   10.08 &  -28.31 &  -25.56 &  -21.50 &    2.758 &    6.817 & 3.5508 & 13\\
IRAC\_4.5 &    3.28 &    6.51 &   11.08 &  -28.29 &  -25.06 &  -20.49 &    3.232 &    7.804 & 4.4960 & 13\\
IRAC\_5.8 &    3.28 &    7.00 &   12.09 &  -28.30 &  -24.58 &  -19.48 &    3.720 &    8.816 & 5.7245 & 13\\
IRAC\_8.0 &    3.26 &    7.62 &   13.41 &  -28.31 &  -23.95 &  -18.16 &    4.360 &   10.152 & 7.8842 & 13\\
IRS\_16 &    3.27 &    9.11 &   16.42 &  -28.31 &  -22.47 &  -15.15 &    5.839 &   13.157 & 15.9222 & 13\\
IRS\_22 &    3.27 &    9.86 &   17.92 &  -28.30 &  -21.72 &  -13.65 &    6.584 &   14.650 & 22.4704 & 14\\
MIPS\_24 &    3.27 &   10.01 &   18.19 &  -28.30 &  -21.57 &  -13.38 &    6.731 &   14.918 & 23.7592 & 14\\
MIPS\_70 &    3.29 &   12.40 &   23.00 &  -28.28 &  -19.17 &   -8.58 &    9.114 &   19.708 & 71.9861 & 14\\
MIPS\_160 &    3.29 &   14.15 &   26.42 &  -28.28 &  -17.43 &   -5.15 &   10.857 &   23.137 & 156.4274 & 14\\
ACS\_F330W &    5.34 &    6.43 &    5.47 &  -26.24 &  -25.14 &  -26.10 &    1.097 &    0.139 & 0.3521 & 14\\
ACS\_F410W &    5.70 &    5.67 &    5.02 &  -25.87 &  -25.90 &  -26.55 &   -0.033 &   -0.680 & 0.4064 & 14\\
ACS\_F435W &    5.48 &    5.35 &    4.84 &  -26.09 &  -26.22 &  -26.73 &   -0.129 &   -0.639 & 0.4328 & 14\\
ACS\_F475W &    5.21 &    5.09 &    4.78 &  -26.36 &  -26.49 &  -26.80 &   -0.122 &   -0.432 & 0.4747 & 14\\
ACS\_F555W &    4.87 &    4.84 &    4.79 &  -26.71 &  -26.74 &  -26.78 &   -0.030 &   -0.076 & 0.5361 & 14\\
ACS\_F606W &    4.66 &    4.72 &    4.89 &  -26.92 &  -26.85 &  -26.68 &    0.063 &    0.233 & 0.5922 & 14\\
ACS\_F625W &    4.49 &    4.63 &    4.94 &  -27.08 &  -26.94 &  -26.64 &    0.140 &    0.448 & 0.6312 & 14\\
ACS\_F775W &    4.16 &    4.52 &    5.26 &  -27.42 &  -27.05 &  -26.31 &    0.364 &    1.103 & 0.7694 & 14\\
ACS\_F814W &    4.12 &    4.52 &    5.36 &  -27.46 &  -27.06 &  -26.22 &    0.400 &    1.239 & 0.8059 & 14\\
ACS\_F850LP &    4.01 &    4.50 &    5.59 &  -27.56 &  -27.07 &  -25.98 &    0.494 &    1.577 & 0.9016 & 14\\
WFC3\_F218W &    9.09 &   10.74 &    8.79 &  -22.48 &  -20.83 &  -22.78 &    1.654 &   -0.298 & 0.2229 & 14\\
WFC3\_F225W &    8.51 &   10.13 &    8.32 &  -23.06 &  -21.44 &  -23.25 &    1.625 &   -0.191 & 0.2372 & 14\\
WFC3\_F336W &    5.49 &    6.64 &    5.58 &  -26.09 &  -24.93 &  -25.99 &    1.158 &    0.094 & 0.3355 & 14\\
WFC3\_F390W &    5.66 &    5.85 &    5.12 &  -25.91 &  -25.73 &  -26.45 &    0.187 &   -0.536 & 0.3924 & 14\\
WFC3\_F438W &    5.50 &    5.32 &    4.81 &  -26.07 &  -26.25 &  -26.76 &   -0.178 &   -0.690 & 0.4326 & 14\\
WFC3\_F475W &    5.19 &    5.07 &    4.77 &  -26.38 &  -26.50 &  -26.80 &   -0.122 &   -0.419 & 0.4774 & 14\\
WFC3\_F555W &    4.91 &    4.86 &    4.79 &  -26.67 &  -26.72 &  -26.78 &   -0.048 &   -0.116 & 0.5308 & 14\\
WFC3\_F606W &    4.67 &    4.73 &    4.88 &  -26.91 &  -26.85 &  -26.69 &    0.059 &    0.217 & 0.5887 & 14\\
WFC3\_F625W &    4.52 &    4.64 &    4.92 &  -27.06 &  -26.93 &  -26.65 &    0.124 &    0.409 & 0.6241 & 14\\
WFC3\_F775W &    4.16 &    4.52 &    5.25 &  -27.41 &  -27.05 &  -26.33 &    0.357 &    1.083 & 0.7648 & 14\\
WFC3\_F814W &    4.12 &    4.52 &    5.35 &  -27.45 &  -27.06 &  -26.22 &    0.395 &    1.226 & 0.8030 & 14\\
WFC3\_F098m &    3.96 &    4.50 &    5.78 &  -27.61 &  -27.07 &  -25.79 &    0.538 &    1.816 & 0.9864 & 14\\
WFC3\_F105W &    3.89 &    4.51 &    5.94 &  -27.68 &  -27.06 &  -25.63 &    0.622 &    2.046 & 1.0551 & 14\\
WFC3\_F125W &    3.66 &    4.54 &    6.33 &  -27.91 &  -27.03 &  -25.24 &    0.877 &    2.667 & 1.2486 & 14\\
WFC3\_F140W &    3.51 &    4.56 &    6.59 &  -28.06 &  -27.01 &  -24.98 &    1.052 &    3.079 & 1.3922 & 14\\
WFC3\_F160W &    3.37 &    4.60 &    6.84 &  -28.20 &  -26.97 &  -24.73 &    1.228 &    3.469 & 1.5370 & 14\\
WFPC2\_F218W &    9.17 &   10.83 &    8.86 &  -22.40 &  -20.74 &  -22.72 &    1.657 &   -0.316 & 0.2207 & 15\\
WFPC2\_F300W &    6.10 &    7.40 &    6.09 &  -25.48 &  -24.17 &  -25.48 &    1.307 &   -0.005 & 0.2992 & 15\\
WFPC2\_F450W &    5.31 &    5.20 &    4.80 &  -26.26 &  -26.37 &  -26.77 &   -0.110 &   -0.509 & 0.4556 & 15\\
WFPC2\_F555W &    4.84 &    4.82 &    4.81 &  -26.73 &  -26.75 &  -26.77 &   -0.025 &   -0.038 & 0.5442 & 15\\
WFPC2\_F606W &    4.62 &    4.70 &    4.90 &  -26.95 &  -26.87 &  -26.67 &    0.077 &    0.276 & 0.6001 & 15\\
WFPC2\_F702W &    4.33 &    4.57 &    5.08 &  -27.24 &  -27.00 &  -26.49 &    0.240 &    0.748 & 0.6919 & 15\\
WFPC2\_F814W &    4.12 &    4.52 &    5.34 &  -27.45 &  -27.05 &  -26.23 &    0.392 &    1.216 & 0.8002 & 15\\
NIC2\_F110W &    3.82 &    4.52 &    6.08 &  -27.75 &  -27.05 &  -25.49 &    0.704 &    2.265 & 1.1235 & 15\\
NIC2\_F160W &    3.35 &    4.64 &    6.97 &  -28.22 &  -26.93 &  -24.60 &    1.286 &    3.618 & 1.6030 & 15\\
NIC3\_F110W &    3.82 &    4.52 &    6.08 &  -27.75 &  -27.05 &  -25.50 &    0.701 &    2.255 & 1.1200 & 15\\
NIC3\_F160W &    3.35 &    4.64 &    6.97 &  -28.22 &  -26.93 &  -24.60 &    1.287 &    3.621 & 1.6042 & 15\\
NIRCAM\_F070W &    4.29 &    4.56 &    5.10 &  -27.28 &  -27.02 &  -26.47 &    0.264 &    0.811 & 0.7046 & 15\\
NIRCAM\_F090W &    4.02 &    4.50 &    5.59 &  -27.56 &  -27.07 &  -25.98 &    0.488 &    1.573 & 0.9025 & 15\\
NIRCAM\_F115W &    3.77 &    4.53 &    6.15 &  -27.80 &  -27.05 &  -25.43 &    0.753 &    2.373 & 1.1543 & 15\\
NIRCAM\_F140M &    3.48 &    4.56 &    6.60 &  -28.09 &  -27.02 &  -24.97 &    1.079 &    3.126 & 1.4053 & 15\\
NIRCAM\_F150W &    3.41 &    4.59 &    6.78 &  -28.16 &  -26.98 &  -24.79 &    1.182 &    3.371 & 1.5007 & 15\\
NIRCAM\_F150W2 &    3.50 &    4.70 &    7.11 &  -28.07 &  -26.87 &  -24.46 &    1.203 &    3.610 & 1.6588 & 15\\
NIRCAM\_F162M &    3.32 &    4.65 &    7.01 &  -28.25 &  -26.93 &  -24.56 &    1.328 &    3.693 & 1.6272 & 15\\
NIRCAM\_F164N &    3.29 &    4.66 &    7.05 &  -28.28 &  -26.91 &  -24.53 &    1.368 &    3.756 & 1.6445 & 15\\
NIRCAM\_F182M &    3.28 &    4.81 &    7.45 &  -28.29 &  -26.76 &  -24.12 &    1.534 &    4.172 & 1.8452 & 15\\
NIRCAM\_F187N &    3.25 &    4.85 &    7.52 &  -28.33 &  -26.72 &  -24.05 &    1.600 &    4.272 & 1.8739 & 15\\
NIRCAM\_F200W &    3.28 &    4.93 &    7.73 &  -28.30 &  -26.64 &  -23.84 &    1.652 &    4.453 & 1.9886 & 15\\
NIRCAM\_F200W &    3.28 &    4.93 &    7.73 &  -28.30 &  -26.64 &  -23.84 &    1.652 &    4.453 & 1.9886 & 15\\
NIRCAM\_F210M &    3.27 &    5.03 &    7.94 &  -28.30 &  -26.54 &  -23.63 &    1.757 &    4.671 & 2.0955 & 15\\
NIRCAM\_F250M &    3.27 &    5.37 &    8.67 &  -28.30 &  -26.21 &  -22.91 &    2.093 &    5.393 & 2.5032 & 15\\
NIRCAM\_F277W &    3.26 &    5.53 &    9.04 &  -28.31 &  -26.04 &  -22.53 &    2.265 &    5.779 & 2.7618 & 15\\
NIRCAM\_F300M &    3.26 &    5.69 &    9.37 &  -28.31 &  -25.88 &  -22.20 &    2.429 &    6.115 & 2.9892 & 15\\
NIRCAM\_F322W2 &    3.26 &    5.77 &    9.63 &  -28.31 &  -25.80 &  -21.95 &    2.509 &    6.365 & 3.2320 & 15\\
NIRCAM\_F323N &    3.26 &    5.84 &    9.70 &  -28.31 &  -25.73 &  -21.87 &    2.583 &    6.441 & 3.2369 & 15\\
NIRCAM\_F335M &    3.26 &    5.92 &    9.86 &  -28.31 &  -25.66 &  -21.71 &    2.658 &    6.599 & 3.3621 & 15\\
NIRCAM\_F356W &    3.26 &    6.02 &   10.09 &  -28.31 &  -25.55 &  -21.48 &    2.763 &    6.833 & 3.5684 & 15\\
NIRCAM\_F405N &    3.24 &    6.30 &   10.65 &  -28.33 &  -25.27 &  -20.93 &    3.058 &    7.404 & 4.0517 & 15\\
NIRCAM\_F410M &    3.26 &    6.31 &   10.67 &  -28.32 &  -25.27 &  -20.90 &    3.049 &    7.411 & 4.0822 & 15\\
NIRCAM\_F430M &    3.27 &    6.41 &   10.88 &  -28.31 &  -25.16 &  -20.69 &    3.147 &    7.613 & 4.2813 & 15\\
NIRCAM\_F444W &    3.27 &    6.46 &   10.99 &  -28.30 &  -25.11 &  -20.59 &    3.185 &    7.712 & 4.4040 & 15\\
NIRCAM\_F460M &    3.29 &    6.60 &   11.23 &  -28.28 &  -24.97 &  -20.34 &    3.308 &    7.943 & 4.6285 & 15\\
NIRCAM\_F466N &    3.26 &    6.62 &   11.26 &  -28.31 &  -24.96 &  -20.31 &    3.352 &    8.000 & 4.6544 & 15\\
NIRCAM\_F470N &    3.29 &    6.63 &   11.30 &  -28.28 &  -24.94 &  -20.27 &    3.341 &    8.013 & 4.7078 & 15\\
NIRCAM\_F480M &    3.29 &    6.67 &   11.39 &  -28.28 &  -24.90 &  -20.18 &    3.383 &    8.104 & 4.8167 & 15\\
MIRI\_F560W &    3.28 &    6.97 &   12.03 &  -28.29 &  -24.60 &  -19.54 &    3.693 &    8.756 & 5.6362 & 16\\
MIRI\_F770W &    3.26 &    7.58 &   13.30 &  -28.31 &  -24.00 &  -18.27 &    4.314 &   10.039 & 7.6428 & 16\\
MIRI\_F1000W &    3.26 &    8.15 &   14.45 &  -28.31 &  -23.42 &  -17.13 &    4.883 &   11.181 & 9.9544 & 16\\
MIRI\_F1130W &    3.26 &    8.43 &   15.00 &  -28.31 &  -23.14 &  -16.57 &    5.166 &   11.741 & 11.3087 & 16\\
MIRI\_F1500W &    3.27 &    9.03 &   16.23 &  -28.31 &  -22.54 &  -15.35 &    5.763 &   12.961 & 15.0651 & 16\\
MIRI\_F1800W &    3.27 &    9.42 &   17.00 &  -28.30 &  -22.15 &  -14.57 &    6.149 &   13.732 & 17.9865 & 16\\
MIRI\_F2100W &    3.27 &    9.72 &   17.62 &  -28.30 &  -21.85 &  -13.95 &    6.453 &   14.351 & 20.7950 & 16\\
MIRI\_F2550W &    3.28 &   10.16 &   18.49 &  -28.30 &  -21.41 &  -13.08 &    6.887 &   15.216 & 25.3639 & 16\\
NIRISS\_F090W &    4.02 &    4.50 &    5.59 &  -27.56 &  -27.07 &  -25.98 &    0.488 &    1.575 & 0.9031 & 17\\
NIRISS\_F115W &    3.78 &    4.53 &    6.14 &  -27.79 &  -27.05 &  -25.43 &    0.747 &    2.358 & 1.1499 & 17\\
NIRISS\_F140M &    3.48 &    4.56 &    6.60 &  -28.09 &  -27.02 &  -24.97 &    1.078 &    3.123 & 1.4044 & 17\\
NIRISS\_F150W &    3.41 &    4.59 &    6.77 &  -28.16 &  -26.98 &  -24.81 &    1.173 &    3.352 & 1.4936 & 17\\
NIRISS\_F158M &    3.35 &    4.62 &    6.93 &  -28.23 &  -26.95 &  -24.64 &    1.277 &    3.582 & 1.5825 & 17\\
NIRISS\_F200W &    3.28 &    4.93 &    7.74 &  -28.30 &  -26.64 &  -23.83 &    1.656 &    4.461 & 1.9930 & 17\\
NIRISS\_F277W &    3.27 &    5.53 &    9.04 &  -28.30 &  -26.05 &  -22.53 &    2.258 &    5.774 & 2.7641 & 17\\
NIRISS\_F356W &    3.26 &    6.03 &   10.11 &  -28.31 &  -25.54 &  -21.46 &    2.769 &    6.854 & 3.5926 & 17\\
NIRISS\_F380M &    3.26 &    6.17 &   10.39 &  -28.31 &  -25.40 &  -21.18 &    2.908 &    7.128 & 3.8229 & 17\\
NIRISS\_F430M &    3.27 &    6.40 &   10.87 &  -28.30 &  -25.17 &  -20.70 &    3.130 &    7.595 & 4.2792 & 17\\
NIRISS\_F444W &    3.27 &    6.47 &   11.00 &  -28.30 &  -25.11 &  -20.57 &    3.191 &    7.729 & 4.4270 & 17\\
NIRISS\_F480M &    3.29 &    6.66 &   11.38 &  -28.28 &  -24.91 &  -20.19 &    3.366 &    8.086 & 4.8113 & 17\\
OMEGACAM\_u &    5.46 &    6.34 &    5.43 &  -26.11 &  -25.23 &  -26.15 &    0.881 &   -0.035 & 0.3590 & 18\\
OMEGACAM\_g &    5.21 &    5.09 &    4.77 &  -26.36 &  -26.48 &  -26.80 &   -0.126 &   -0.442 & 0.4735 & 18\\
OMEGACAM\_r &    4.50 &    4.63 &    4.93 &  -27.07 &  -26.94 &  -26.64 &    0.133 &    0.429 & 0.6276 & 18\\
OMEGACAM\_i &    4.20 &    4.53 &    5.21 &  -27.38 &  -27.05 &  -26.36 &    0.331 &    1.013 & 0.7495 & 18\\
OMEGACAM\_z &    4.01 &    4.51 &    5.55 &  -27.56 &  -27.07 &  -26.03 &    0.493 &    1.534 & 0.8842 & 18\\
VIRCAM\_Z &    4.02 &    4.51 &    5.56 &  -27.56 &  -27.07 &  -26.01 &    0.491 &    1.546 & 0.8899 & 19\\
VIRCAM\_Y &    3.93 &    4.51 &    5.87 &  -27.64 &  -27.07 &  -25.70 &    0.577 &    1.940 & 1.0253 & 19\\
VIRCAM\_H &    3.65 &    4.54 &    6.34 &  -27.93 &  -27.03 &  -25.23 &    0.892 &    2.691 & 1.2535 & 19\\
VIRCAM\_J &    3.32 &    4.66 &    7.05 &  -28.25 &  -26.91 &  -24.53 &    1.335 &    3.721 & 1.6430 & 19\\
VIRCAM\_Ks &    3.27 &    5.07 &    8.04 &  -28.30 &  -26.50 &  -23.53 &    1.797 &    4.767 & 2.1494 & 19\\
SkyMapper\_u &    5.33 &    6.32 &    5.40 &  -26.24 &  -25.25 &  -26.17 &    0.989 &    0.073 & 0.3590 & 20\\
SkyMapper\_v &    5.81 &    6.09 &    5.31 &  -25.77 &  -25.49 &  -26.26 &    0.280 &   -0.493 & 0.3836 & 20\\
SkyMapper\_g &    5.03 &    4.94 &    4.78 &  -26.55 &  -26.63 &  -26.79 &   -0.082 &   -0.247 & 0.5075 & 20\\
SkyMapper\_r &    4.56 &    4.66 &    4.91 &  -27.02 &  -26.91 &  -26.66 &    0.104 &    0.352 & 0.6138 & 20\\
SkyMapper\_i &    4.14 &    4.52 &    5.28 &  -27.43 &  -27.05 &  -26.29 &    0.377 &    1.137 & 0.7768 & 20\\
SkyMapper\_z &    4.00 &    4.50 &    5.62 &  -27.57 &  -27.07 &  -25.95 &    0.502 &    1.615 & 0.9143 & 20\\
\enddata
\tablerefs{
  1: \citet{2015PASP..127..102M}; 
  2: \citet{Cohen2mass};                         
  3: \citet{SDSS};                               
  4: \cite{DECAM};                               
  5: \citet{Tonry2012};                          
  6: \cite{Gwyn2012};                            
  7: Nick Kaiser 2002, private communication;   
  8: \citet{Hewett2006};                         
  9: \href{https://github.com/lsst/throughputs/tree/master/baseline}{https://github.com/lsst/throughputs/tree/master/baseline}
  10: \citet{BM12};
  11: \citet{BessellBrett};                      
  12: \citet{Galex};                             
  13: \citet{JarrettWise};                       
  14: \citet{IRAS};                              
  15: \citet{Spitzer};                           
  15: \citet{calspec};                           
  16: \citet{jwst-docs}.                        
  17: \citet{jwst-docs}.  
  18: \href{https://www.eso.org/sci/facilities/paranal/instruments/omegacam/tools.html}{https://www.eso.org/sci/facilities/paranal/instruments/omegacam/tools.html}
  19: \href{https://www.eso.org/sci/facilities/paranal/instruments/vircam/inst.html}{https://www.eso.org/sci/facilities/paranal/instruments/vircam/inst.html}
  20: \citet{Bessell2011}.
}
\end{deluxetable*}  


\section{Summary}

This work uses the dust-free composite spectrum of Vega 
with the absolute calibration set by Sirius, both due
to \citet{2014AJ....147..127B}, to calculate a table with the absolute
magnitude of the Sun and the conversion between the $vegamag$ and the
AB and ST systems for
several filters used in ground and space-based observatories. The solar
SED used in this paper is a composite combining space-based
spectra of the Sun from the ultra-violet to the near-infrared
\citep{2017JGRA..122.5910H}, with models of the solar atmosphere out
to 300 $\mu$m \citep{Fontenla2011, Kurucz2011}.
For the set of Johnson ($U$, $B$, $V$) and Cousins ($R$ and $I$)
filters, which are originally characterized using photoelectric
photometry, filter curves reconstructed using Monte Carlo methods by
\citet{2015PASP..127..102M} are used. To verify the consistency of the
synthetic spectra measured using the composite spectra of Vega and the
Sun, the colors measured for these SEDs are compared with photometric
measurements of AV stellar templates and solar analogs respectively. 
The comparison between colors calculated for the Vega SED
and AV stars shows absolute offsets $<$ 0.01 magnitudes and a
dispersion $<$ 0.03 magnitudes, consistent with the estimated
uncertainty at the 2\% level for the Vega SED by \citet{2014AJ....147..127B}.
The comparison between colors measured with the solar composite and
the solar analogs of \citet{Ramirez2012} and \citet{Casagrande2012} shows an
offset of $\sim$ -0.02 $\pm$ 0.03 magnitudes. Assuming the errors are
equally distributed, this translates to an average uncertainty of
$\sim$ 2\% for the solar SED. Adding in quadrature the uncertainty
in the calibration of both spectra translates to errors 
$\sim$~3--4\% for the solar absolute magnitudes.

\appendix

\section{Filter parameters}

As shown by \citet{Rieke2008} and BM12, there are a number of definitions
used to characterize filter properties and frequently the names
associated to these are inconsistent in the literature (BM12).
For convenience the expressions used to calculate the
filter parameters are presented here and
the reader is referred to Appendix E of \citet{Rieke2008},
\replaced{and the Appendix of \citet{BM12} }{
the Appendix of \citet{BM12} 
and the review of \citet{Bohlinetal2014}
}
for more detailed discussions on  the determination,
history and naming of these definitions.

The following characteristic wavelengths are only dependent on the
filter shape. The {\it{mean photon wavelength}} \citep{BM12}, also
called {\it{mean wavelength}} by \citep{Tokunaga2005} and $mean$ or
{\it{effective wavelength}} \citet{Rieke2008} is defined as

\begin{equation}
\lambda_{mean} = \frac{\int R(\lambda)\lambda d\lambda}{\int R(\lambda) d\lambda}
\end{equation}

The mean flux of a source within the band is defined as
\begin{equation}
\langle f_{\lambda} \rangle = \frac{\int f_\lambda R(\lambda) \lambda
  d\lambda}{\int R(\lambda)\lambda d\lambda}
\end{equation}

The {\it{nominal wavelength}} of \citet{Rieke2008} is called {\it{mean
energy wavelength}} by \citet{BM12} :
\begin{equation}
\lambda_{n1} = \frac{\int R(\lambda)\lambda^2 d\lambda}{\int R(\lambda)\lambda d\lambda}
\end{equation}

while \citet{Reach2005} define the {\it{nominal wavelength}} as
\begin{equation}
\lambda_{n2} = \frac{\int R(\lambda) d\lambda}{\int \frac {R(\lambda) d\lambda} {\lambda}}
\end{equation}
and in both cases minimize the color correction in a given band
\citep{Reach2005, Rieke2008}.

The {\it{pivot wavelength}}
\begin{equation}
\lambda_{pivot} = \sqrt \frac{\int R(\lambda)\lambda
  d\lambda}{\int \frac {R(\lambda) d\lambda} {\lambda}}
\end{equation}
is the wavelength where
$\langle f_{\lambda} \rangle \frac {\lambda_{pivot}^2}{c} = \langle f_{\nu} \rangle$, and
$\langle f_{\lambda}\rangle$ or $\langle f_{\nu} \rangle$ are the mean
flux density within the band.

The following characteristic wavelengths also take into account the flux
density of the source ($f_\lambda$). As noted by BM12
there are a multiplicity of definitions for the {\it{effective
wavelength}} and they propose this as the standard: 
\begin{equation}
\lambda_{eff} = \frac{\int f_\lambda(\lambda) R(\lambda) \lambda^2
  d\lambda}{\int Rf_\lambda(\lambda) (\lambda)\lambda d\lambda}
\end{equation}

The wavelength where the monochromatic flux of a source
is equivalent to the average flux of the source within the band
is defined as the {\it{isophotal wavelength}} \citep{Cohen1992, Tokunaga2005, Rieke2008}, BM12:
\begin{equation}
f(\lambda_{iso}) = \langle f_{\lambda} \rangle .
\end{equation}

Because this
measurement can be affected by the instrumental resolution and the
presence of stellar lines \citep{Rieke2008}, when calculating the
isophotal wavelength one may need to  smooth the spectrum prior to the
calculation \citep{BM12}, use a continuum model or interpolate over
spectral lines \citep{Rieke2008, Cohen1992}.

The {\it{bandwidth}} is defined as the integral of the normalized
transmission \citep{Budding1993}, and the following definition is
adopted by \citet{Rieke2008} and \citet{2015PASP..127..102M} (where it
is called {\it{effective width}}):
\begin{equation}
BW = \frac{\int R(\lambda) d\lambda}{max[R(\lambda)]}
\end{equation}

The {\it{average system response}} is 
\begin{equation}
Resp = \frac{\int R(\lambda) d\lambda}{\int d\lambda}
\end{equation}

For NIRCam filters tabulated in \citet{jwst-docs}, an {\it{effective
response}} is adopted where
\begin{equation}
R_{eff} = \frac
{\int_{\lambda\_{pivot}-BW/2}^{\lambda\_{pivot}+BW/2}R(\lambda)
  d\lambda} {BW}
\end{equation}

The $vegamag$ zero-point is defined as
\begin{equation}
zp = +2.5~log_{10}~[\frac {\int f_{\lambda}(\lambda) R(\lambda)\lambda
    d\lambda} {\int R(\lambda)\lambda d\lambda}]
\end{equation}
while the flux at zero magnitude is calculated by using the spectrum of Vega
(corrected to have zero magnitude in all bands)
such that 
\begin{equation}
f{_\nu0} = \frac { \lambda_{pivot}^{2}} {c} \frac {\int f_{\lambda}(\lambda) R(\lambda)\lambda d\lambda} {\int R(\lambda)\lambda d\lambda}
\end{equation}
converted into Jansky, where $c$ is the speed of light.
Table \ref{tab:2} shows these parameters calculated for the filters
used in Table \ref{tab:1}, where the BM12 naming is used. The
table lists in column (1) the filter name, 
in column (2) the mean photon wavelength,
column (3) the pivot wavelength,
column (4) the effective wavelength 
column (5) the nominal wavelength using the \citet{Rieke2008}
definition; 
column (6) the nominal wavelength using the \citet{Reach2005}
definition; 
the isophotal wavelength using the \citet{Rieke2008} column (7) and
BM12 column (8) definitions. Column 9 shows the wavelength range,
column (10) the bandwidth, column (11) the full width at half-maximum,
column (12) the filter response. The zero-point for $vegamag$ follows
in column (14), the corresponding flux density in erg $s^{-1}$
$cm^{-2}$ $\AA^{-1}$, and in Jansky in column (15).

\acknowledgments
I thank George Rieke for suggestions and discussions. I also thank the
anonymous referee and the editor Dr. Shadia Habbal whose suggestions
helped improving the presentation.
Funding from the JWST/NIRCam contract NAS5-02015 to the University of
Arizona,
the use of the NASA/SAO ADS, the NASA/IPAC Infrared Science Archive,
Simbad (at Strasbourg and Harvard) and the Mikulski Archive for Space
Telescopes are  gratefully acknowledged.
This work uses data from the SOLID Project
\href{http://projects.pmodwrc.ch/solid/}{http://projects.pmodwrc.ch/solid/} which is funded by the 
European Community's Seventh Framework Programne (FP7 2012) under
grant agreement no 313188.


\facilities{IRSA, STScI:CDBS, STScI:MAST, JWST-Docs, GALEX, ADS, Simbad}

\clearpage
\begin{longrotatetable}
\begin{deluxetable}{lrrrrrrrrrrrrcr}
\tabletypesize {\tiny}
\tablecaption{Filter Parameters\label{tab:2}}
\tablehead{
\colhead{Filter} & \colhead{$\lambda_{mean}$} & \colhead{$\lambda_{pivot}$} & \colhead{$\lambda_{eff}$} & \colhead{$\lambda_{n1}$} & \colhead{$\lambda_{n2}$} & \colhead{$\lambda_i$} & \colhead{$\lambda_i$(BM12)} & \colhead{$\lambda_{range}$} & \colhead{BW} & \colhead{FWHM} & \colhead{response} & \colhead{$zp$} & \colhead{f$_{\lambda}(zp)$} & \colhead{f$_{\nu}$(mag0)}\\
 {}       & $\mu$m~~~~ & $\mu$m~~~~ & $\mu$m~~~~ & $\mu$m~~~~ & $\mu$m~~~~ & $\mu$m~~~~ &
 $\mu$m~~~~ & $\mu$m~~~~ & $\mu$m~~~~ & $\mu$m~~~~ & {} & {}  & erg  s$^{-1}$ cm$^{-2}$ \AA$^{-1}$ & Jy~~~~~ \\
(1) & (2)~~~~ & (3)~~~~ & (4)~~~~ & (5)~~~~ & (6)~~~~ & (7)~~~~ & (8)~~~~ & (9)~~~~ & (10)~~~~ & (11)~~~~ & (12)~~~~ & (13)~~ & (14) & (15)~~~~\\}
\startdata
Johnson\_U & 0.3618 & 0.3611 & 0.3694 & 0.3633 & 0.3603 & 0.3691 & 0.3719 & 0.1238 & 0.0581 & 0.0561 & 0.4693 & 20.9170 & 4.29723E-09 & 1868.72\\
Johnson\_B & 0.4410 & 0.4396 & 0.4390 & 0.4438 & 0.4382 & 0.3896 & 0.3911 & 0.1856 & 0.0992 & 0.1004 & 0.5345 & 20.4951 & 6.33795E-09 & 4085.60\\
Johnson\_V & 0.5524 & 0.5511 & 0.5476 & 0.5551 & 0.5499 & 0.5527 & 0.5510 & 0.2657 & 0.0871 & 0.0812 & 0.3275 & 21.1011 & 3.62701E-09 & 3674.73\\
Cousins\_R & 0.6612 & 0.6582 & 0.6492 & 0.6674 & 0.6553 & 0.6545 & 0.6545 & 0.3322 & 0.1669 & 0.1671 & 0.5024 & 21.6781 & 2.13191E-09 & 3080.98\\
Cousins\_I & 0.8047 & 0.8034 & 0.7993 & 0.8074 & 0.8020 & 0.8006 & 0.8008 & 0.2163 & 0.1482 & 0.1523 & 0.6850 & 22.3469 & 1.15139E-09 & 2478.76\\
Tycho\_Bt & 0.4220 & 0.4212 & 0.4234 & 0.4237 & 0.4204 & 0.3944 & 0.3938 & 0.1548 & 0.0741 & 0.0739 & 0.4790 & 20.4332 & 6.70971E-09 & 3970.61\\
Tycho\_Vt & 0.5352 & 0.5335 & 0.5291 & 0.5389 & 0.5317 & 0.5359 & 0.5316 & 0.2263 & 0.1134 & 0.1116 & 0.5009 & 21.0088 & 3.94906E-09 & 3748.93\\
Hipparcos\_Hp & 0.5596 & 0.5508 & 0.5315 & 0.5780 & 0.5421 & 0.5467 & 0.5522 & 0.5772 & 0.2405 & 0.2237 & 0.4166 & 21.1107 & 3.59516E-09 & 3638.04\\
2MASS\_J & 1.2411 & 1.2393 & 1.2321 & 1.2445 & 1.2376 & 1.2378 & 1.2377 & 0.3523 & 0.1628 & 0.2027 & 0.4611 & 23.7442 & 3.17931E-10 & 1628.84\\
2MASS\_H & 1.6513 & 1.6495 & 1.6424 & 1.6551 & 1.6476 & 1.6467 & 1.6467 & 0.3826 & 0.2510 & 0.2610 & 0.6559 & 24.8385 & 1.16041E-10 & 1053.12\\
2MASS\_Ks & 2.1656 & 2.1638 & 2.1558 & 2.1691 & 2.1621 & 2.1622 & 2.1625 & 0.4401 & 0.2622 & 0.2785 & 0.5951 & 25.8980 & 4.37335E-11 & 683.04\\
SDSS\_u & 0.3562 & 0.3556 & 0.3607 & 0.3572 & 0.3551 & 0.3642 & 0.3663 & 0.0983 & 0.0558 & 0.0582 & 0.0616 & 21.0632 & 3.75600E-09 & 1584.71\\
SDSS\_g & 0.4719 & 0.4702 & 0.4673 & 0.4751 & 0.4686 & 0.4730 & 0.4747 & 0.1963 & 0.1158 & 0.1263 & 0.2132 & 20.6442 & 5.52461E-09 & 4075.09\\
SDSS\_r & 0.6185 & 0.6176 & 0.6142 & 0.6204 & 0.6166 & 0.6169 & 0.6169 & 0.1723 & 0.1111 & 0.1150 & 0.3170 & 21.4800 & 2.55856E-09 & 3254.86\\
SDSS\_i & 0.7500 & 0.7490 & 0.7459 & 0.7519 & 0.7480 & 0.7498 & 0.7483 & 0.1844 & 0.1045 & 0.0683 & 0.2397 & 22.1124 & 1.42903E-09 & 2674.11\\
SDSS\_z & 0.8961 & 0.8947 & 0.8925 & 0.8992 & 0.8932 & 0.8933 & 0.8971 & 0.2781 & 0.1125 & 0.0994 & 0.0318 & 22.6604 & 8.62665E-10 & 2303.28\\
DES\_u & 0.3879 & 0.3859 & 0.3881 & 0.3970 & 0.3839 & 0.3774 & 0.3805 & 0.0711 & 0.0278 & 0.0256 & 0.0551 & 20.6479 & 5.50603E-09 & 2735.17\\
DES\_g & 0.4842 & 0.4820 & 0.4776 & 0.4890 & 0.4798 & 0.3811 & 0.3794 & 0.1663 & 0.1141 & 0.1299 & 0.2573 & 20.7091 & 5.20407E-09 & 4033.18\\
DES\_r & 0.6439 & 0.6423 & 0.6374 & 0.6470 & 0.6408 & 0.6379 & 0.6373 & 0.1901 & 0.1383 & 0.1484 & 0.3826 & 21.6052 & 2.27988E-09 & 3137.37\\
DES\_i & 0.7821 & 0.7807 & 0.7758 & 0.7848 & 0.7792 & 0.7626 & 0.7792 & 0.4647 & 0.1393 & 0.1482 & 0.1696 & 22.2523 & 1.25624E-09 & 2553.68\\
DES\_z & 0.9172 & 0.9158 & 0.9139 & 0.9196 & 0.9145 & 0.9574 & 0.9235 & 0.6876 & 0.1270 & 0.1479 & 0.1043 & 22.7104 & 8.23857E-10 & 2304.99\\
DES\_Y & 0.9877 & 0.9866 & 0.9830 & 0.9893 & 0.9855 & 0.9979 & 0.9898 & 0.1830 & 0.0680 & 0.0664 & 0.1642 & 22.9192 & 6.79718E-10 & 2207.09\\
PS1\_g & 0.4866 & 0.4849 & 0.4811 & 0.4900 & 0.4832 & 0.4866 & 0.4879 & 0.1707 & 0.1166 & 0.1256 & 0.3430 & 20.7241 & 5.13274E-09 & 4025.81\\
PS1\_r & 0.6215 & 0.6201 & 0.6156 & 0.6241 & 0.6188 & 0.6195 & 0.6200 & 0.1768 & 0.1318 & 0.1404 & 0.5121 & 21.4901 & 2.53499E-09 & 3251.66\\
PS1\_i & 0.7545 & 0.7535 & 0.7504 & 0.7564 & 0.7525 & 0.7525 & 0.7531 & 0.1659 & 0.1243 & 0.0698 & 0.6509 & 22.1328 & 1.40245E-09 & 2656.00\\
PS1\_z & 0.8679 & 0.8674 & 0.8669 & 0.8690 & 0.8669 & 0.8597 & 0.8662 & 0.1519 & 0.0966 & 0.1034 & 0.5596 & 22.5824 & 9.26899E-10 & 2326.30\\
PS1\_Y & 0.9633 & 0.9628 & 0.9614 & 0.9645 & 0.9622 & 0.9645 & 0.9621 & 0.1997 & 0.0616 & 0.0629 & 0.1893 & 22.8407 & 7.30650E-10 & 2259.11\\
cfhtls\_u & 0.3811 & 0.3803 & 0.3895 & 0.3829 & 0.3794 & 0.3810 & 0.3813 & 0.2267 & 0.0575 & 0.0654 & 0.1078 & 20.6447 & 5.52249E-09 & 2663.85\\
cfhtls\_g & 0.4862 & 0.4844 & 0.4803 & 0.4899 & 0.4826 & 0.4787 & 0.3774 & 0.2072 & 0.1322 & 0.1434 & 0.4045 & 20.7182 & 5.16084E-09 & 4039.99\\
cfhtls\_r & 0.6258 & 0.6248 & 0.6212 & 0.6279 & 0.6237 & 0.6241 & 0.6237 & 0.2001 & 0.1099 & 0.1219 & 0.3026 & 21.5171 & 2.47266E-09 & 3219.42\\
cfhtls\_i & 0.7690 & 0.7678 & 0.7638 & 0.7715 & 0.7666 & 0.7662 & 0.7667 & 0.2264 & 0.1221 & 0.1367 & 0.2594 & 22.1959 & 1.32325E-09 & 2602.12\\
cfhtls\_z & 0.8870 & 0.8859 & 0.8845 & 0.8894 & 0.8848 & 0.8840 & 0.8856 & 0.2270 & 0.0998 & 0.0936 & 0.1217 & 22.6349 & 8.83167E-10 & 2312.11\\
CFHT\_12kx8k\_B & 0.4407 & 0.4399 & 0.4400 & 0.4424 & 0.4390 & 0.3989 & 0.3879 & 0.2573 & 0.0619 & 0.0605 & 0.2405 & 20.4681 & 6.49800E-09 & 4193.87\\
CFHT\_12kx8k\_R & 0.6621 & 0.6610 & 0.6578 & 0.6642 & 0.6600 & 0.6591 & 0.6581 & 0.1721 & 0.1077 & 0.1181 & 0.6257 & 21.7046 & 2.08038E-09 & 3032.27\\
CFHT\_12kx8k\_I & 0.8183 & 0.8159 & 0.8096 & 0.8231 & 0.8136 & 0.8077 & 0.8107 & 0.2585 & 0.1921 & 0.2139 & 0.7409 & 22.3816 & 1.11520E-09 & 2476.56\\
UKIRT\_z & 0.8831 & 0.8826 & 0.8823 & 0.8840 & 0.8822 & 0.8820 & 0.8812 & 0.1403 & 0.0879 & 0.0926 & 0.1194 & 22.6261 & 8.90324E-10 & 2313.54\\
UKIRT\_Y & 1.0319 & 1.0315 & 1.0299 & 1.0329 & 1.0310 & 1.0307 & 1.0321 & 0.1569 & 0.1008 & 0.1034 & 0.1194 & 23.0663 & 5.93591E-10 & 2106.53\\
UKIRT\_J & 1.2511 & 1.2502 & 1.2462 & 1.2529 & 1.2492 & 1.2476 & 1.2490 & 0.2386 & 0.1475 & 0.1589 & 0.1321 & 23.7842 & 3.06416E-10 & 1597.42\\
UKIRT\_H & 1.6383 & 1.6360 & 1.6271 & 1.6430 & 1.6337 & 1.6313 & 1.6324 & 0.4649 & 0.2773 & 0.2918 & 0.1586 & 24.8055 & 1.19624E-10 & 1067.96\\
UKIRT\_K & 2.2085 & 2.2060 & 2.1950 & 2.2135 & 2.2035 & 2.2017 & 2.2032 & 0.5488 & 0.3276 & 0.3413 & 0.1422 & 25.9737 & 4.07863E-11 & 662.09\\
LSST\_u & 0.3671 & 0.3665 & 0.3743 & 0.3681 & 0.3660 & 0.3748 & 0.3724 & 0.0906 & 0.0547 & 0.0623 & 0.0829 & 20.8555 & 4.54799E-09 & 2038.03\\
LSST\_g & 0.4827 & 0.4808 & 0.4768 & 0.4864 & 0.4789 & 0.4841 & 0.4859 & 0.1799 & 0.1333 & 0.1426 & 0.3027 & 20.7014 & 5.24138E-09 & 4041.31\\
LSST\_r & 0.6223 & 0.6210 & 0.6165 & 0.6250 & 0.6197 & 0.6206 & 0.6210 & 0.1685 & 0.1338 & 0.1343 & 0.3605 & 21.4946 & 2.52446E-09 & 3247.24\\
LSST\_i & 0.7546 & 0.7537 & 0.7506 & 0.7565 & 0.7527 & 0.7527 & 0.7533 & 0.1565 & 0.1209 & 0.0680 & 0.3490 & 22.1336 & 1.40133E-09 & 2655.04\\
LSST\_z & 0.8691 & 0.8686 & 0.8680 & 0.8702 & 0.8680 & 0.8622 & 0.8681 & 0.1350 & 0.0994 & 0.1022 & 0.3269 & 22.5857 & 9.24140E-10 & 2325.47\\
LSST\_y & 0.9710 & 0.9705 & 0.9688 & 0.9722 & 0.9699 & 0.9686 & 0.9717 & 0.1828 & 0.0814 & 0.0857 & 0.1259 & 22.8627 & 7.16026E-10 & 2249.34\\
Bessell\_Murphy\_U & 0.3604 & 0.3597 & 0.3674 & 0.3617 & 0.3591 & 0.3674 & 0.3680 & 0.1098 & 0.0621 & 0.0628 & 0.5656 & 20.9560 & 4.14586E-09 & 1789.43\\
Bessell\_Murphy\_B & 0.4391 & 0.4378 & 0.4369 & 0.4420 & 0.4364 & 0.4601 & 0.3914 & 0.1796 & 0.0916 & 0.0894 & 0.5098 & 20.4800 & 6.42715E-09 & 4108.32\\
Bessell\_Murphy\_V & 0.5501 & 0.5489 & 0.5457 & 0.5525 & 0.5477 & 0.5513 & 0.5483 & 0.2495 & 0.0875 & 0.0836 & 0.3507 & 21.0884 & 3.66982E-09 & 3687.87\\
Bessell\_Murphy\_R & 0.6554 & 0.6524 & 0.6436 & 0.6616 & 0.6495 & 0.6483 & 0.6483 & 0.3393 & 0.1485 & 0.1447 & 0.4375 & 21.6485 & 2.19072E-09 & 3110.41\\
Bessell\_Murphy\_I & 0.7996 & 0.7984 & 0.7943 & 0.8023 & 0.7971 & 0.7956 & 0.7960 & 0.1996 & 0.1427 & 0.1498 & 0.7147 & 22.3268 & 1.17294E-09 & 2493.70\\
Bessell\_Murphy\_Bt & 0.4198 & 0.4190 & 0.4215 & 0.4214 & 0.4182 & 0.3928 & 0.3941 & 0.1447 & 0.0719 & 0.0719 & 0.4966 & 20.4308 & 6.72504E-09 & 3938.09\\
Bessell\_Murphy\_Vt & 0.5315 & 0.5300 & 0.5266 & 0.5345 & 0.5285 & 0.5283 & 0.5295 & 0.2096 & 0.0993 & 0.0963 & 0.4737 & 20.9915 & 4.01224E-09 & 3759.32\\
Bessell\_Murphy\_Hp & 0.5429 & 0.5349 & 0.5188 & 0.5595 & 0.5271 & 0.5694 & 0.5352 & 0.5289 & 0.2269 & 0.2117 & 0.4290 & 21.0316 & 3.86674E-09 & 3691.00\\
Bessell\_88\_J & 1.2369 & 1.2347 & 1.2258 & 1.2412 & 1.2325 & 1.2322 & 1.2326 & 0.3593 & 0.2029 & 0.2066 & 0.5308 & 23.7318 & 3.21587E-10 & 1635.23\\
Bessell\_88\_H & 1.6472 & 1.6450 & 1.6365 & 1.6517 & 1.6428 & 1.6406 & 1.6414 & 0.3393 & 0.2845 & 0.2984 & 0.8301 & 24.8263 & 1.17350E-10 & 1059.24\\
Bessell\_88\_K & 2.1683 & 2.1663 & 2.1574 & 2.1721 & 2.1644 & 2.1596 & 2.1637 & 0.3593 & 0.2837 & 0.3048 & 0.7738 & 25.9019 & 4.35755E-11 & 682.12\\
Bessell\_88\_L & 3.4838 & 3.4797 & 3.4602 & 3.4919 & 3.4756 & 3.4728 & 3.4733 & 0.7186 & 0.4583 & 0.5103 & 0.5611 & 27.8367 & 7.33364E-12 & 296.20\\
Bessell\_88\_Lprime & 3.8285 & 3.8247 & 3.8063 & 3.8362 & 3.8208 & 3.8154 & 3.8182 & 0.6786 & 0.5339 & 0.5880 & 0.7632 & 28.2347 & 5.08305E-12 & 248.03\\
Bessell\_88\_M & 4.7369 & 4.7347 & 4.7250 & 4.7411 & 4.7326 & 4.7241 & 4.7325 & 0.5589 & 0.3498 & 0.2044 & 0.3127 & 29.1335 & 2.22134E-12 & 166.11\\
GALEX\_FUV & 0.1539 & 0.1535 & 0.1549 & 0.1546 & 0.1532 & 0.1464 & 0.1469 & 0.0453 & 0.0255 & 0.0228 & 0.0106 & 20.4239 & 6.76768E-09 & 531.97\\
GALEX\_NUV & 0.2316 & 0.2301 & 0.2304 & 0.2345 & 0.2286 & 0.2272 & 0.2269 & 0.1185 & 0.0730 & 0.0796 & 0.0193 & 20.8469 & 4.58413E-09 & 809.45\\
WISE\_1 & 3.4003 & 3.3897 & 3.3387 & 3.4204 & 3.3792 & 3.3687 & 3.3722 & 1.3441 & 0.6628 & 0.6358 & 0.4930 & 27.7140 & 8.21074E-12 & 314.69\\
WISE\_2 & 4.6520 & 4.6406 & 4.5870 & 4.6746 & 4.6293 & 4.6204 & 4.6199 & 1.4623 & 1.0423 & 1.1073 & 0.7128 & 29.0322 & 2.43841E-12 & 175.16\\
WISE\_3 & 12.8114 & 12.5705 & 11.3086 & 13.2371 & 12.3341 & 11.6601 & 12.0626 & 18.3366 & 5.5114 & 6.2771 & 0.3003 & 33.1194 & 5.65225E-14 & 29.79\\
WISE\_4 & 22.3753 & 22.3142 & 22.0230 & 22.5013 & 22.2533 & 22.1724 & 22.1950 & 8.8919 & 4.1023 & 3.6087 & 0.4613 & 35.7525 & 5.00014E-15 & 8.30\\
IRAS12 & 11.5406 & 11.3562 & 10.4650 & 11.8905 & 11.1747 & 10.8564 & 10.9983 & 7.4850 & 5.9671 & 6.9307 & 0.7971 & 32.7211 & 8.15757E-14 & 35.09\\
IRAS25 & 23.8767 & 23.6079 & 22.2580 & 24.3900 & 23.3421 & 23.1021 & 23.0659 & 14.9700 & 10.0234 & 11.2592 & 0.6688 & 35.9191 & 4.28907E-15 & 7.97\\
IRAS60 & 61.4459 & 60.3699 & 54.5695 & 63.3790 & 59.3127 & 52.8094 & 58.0207 & 53.8919 & 30.4317 & 32.7622 & 0.5646 & 39.9327 & 1.06394E-16 & 1.29\\
IRAS100 & 101.9433 & 101.1267 & 96.9972 & 103.5466 & 100.3167 & 99.6179 & 99.4636 & 69.8599 & 33.2387 & 32.2401 & 0.4754 & 42.2860 & 1.21782E-17 & 0.42\\
IRAC\_3.6 & 3.5573 & 3.5508 & 3.5204 & 3.5701 & 3.5443 & 3.5375 & 3.5400 & 0.8893 & 0.6836 & 0.7432 & 0.3639 & 27.9174 & 6.80809E-12 & 286.32\\
IRAC\_4.5 & 4.5049 & 4.4960 & 4.4543 & 4.5228 & 4.4870 & 4.4785 & 4.4786 & 1.3434 & 0.8650 & 1.0097 & 0.3529 & 28.9037 & 2.74485E-12 & 185.07\\
IRAC\_5.8 & 5.7386 & 5.7245 & 5.6564 & 5.7664 & 5.7104 & 5.6999 & 5.6972 & 1.6151 & 1.2562 & 1.3912 & 0.1105 & 29.9163 & 1.08012E-12 & 118.07\\
IRAC\_8.0 & 7.9274 & 7.8842 & 7.6741 & 8.0118 & 7.8413 & 7.7845 & 7.8010 & 3.3582 & 2.5292 & 2.8311 & 0.2365 & 31.2516 & 3.15764E-13 & 65.47\\
IRS\_16 & 16.0478 & 15.9222 & 15.4020 & 16.3590 & 15.7975 & 15.7463 & 15.7090 & 22.8531 & 4.7674 & 5.4763 & 0.6163 & 34.2569 & 1.98254E-14 & 16.77\\
IRS\_22 & 22.6224 & 22.4704 & 21.7563 & 22.9355 & 22.3193 & 22.2729 & 22.1796 & 18.7995 & 7.0115 & 7.3067 & 0.8122 & 35.7499 & 5.01246E-15 & 8.44\\
MIPS\_24 & 23.8436 & 23.7592 & 23.3583 & 24.0181 & 23.6750 & 23.6079 & 23.5923 & 12.6666 & 5.2969 & 5.3248 & 0.4181 & 36.0180 & 3.91555E-15 & 7.37\\
MIPS\_70 & 72.5564 & 71.9861 & 69.3644 & 73.7885 & 71.4202 & 70.8090 & 70.9157 & 60.4363 & 21.3011 & 18.9838 & 0.3527 & 40.8081 & 4.75089E-17 & 0.82\\
MIPS\_160 & 156.9627 & 156.4274 & 153.6888 & 158.0193 & 155.8939 & 155.4756 & 155.3366 & 92.3398 & 35.7629 & 34.5528 & 0.3872 & 44.2365 & 2.02017E-18 & 0.16\\
ACS\_F330W & 0.3522 & 0.3521 & 0.3523 & 0.3525 & 0.3520 & 0.3485 & 0.3593 & 0.0474 & 0.0261 & 0.0272 & 0.0454 & 21.2385 & 3.19592E-09 & 1321.74\\
ACS\_F410W & 0.4069 & 0.4064 & 0.4096 & 0.4078 & 0.4059 & 0.4535 & 0.3954 & 0.0890 & 0.0522 & 0.0543 & 0.2040 & 20.4195 & 6.79505E-09 & 3743.07\\
ACS\_F435W & 0.4338 & 0.4328 & 0.4341 & 0.4358 & 0.4318 & 0.3892 & 0.3940 & 0.1330 & 0.0863 & 0.0935 & 0.2387 & 20.4608 & 6.54181E-09 & 4087.92\\
ACS\_F475W & 0.4766 & 0.4747 & 0.4710 & 0.4802 & 0.4728 & 0.4812 & 0.4805 & 0.1781 & 0.1359 & 0.1437 & 0.2807 & 20.6676 & 5.40703E-09 & 4064.13\\
ACS\_F555W & 0.5373 & 0.5361 & 0.5333 & 0.5398 & 0.5349 & 0.5339 & 0.5329 & 0.1715 & 0.1125 & 0.1240 & 0.2408 & 21.0244 & 3.89261E-09 & 3731.78\\
ACS\_F606W & 0.5960 & 0.5922 & 0.5812 & 0.6035 & 0.5883 & 0.5895 & 0.5917 & 0.2601 & 0.1996 & 0.2323 & 0.3588 & 21.3334 & 2.92845E-09 & 3425.32\\
ACS\_F625W & 0.6325 & 0.6312 & 0.6267 & 0.6352 & 0.6298 & 0.6295 & 0.6307 & 0.1715 & 0.1308 & 0.1416 & 0.3372 & 21.5483 & 2.40256E-09 & 3192.67\\
ACS\_F775W & 0.7707 & 0.7694 & 0.7654 & 0.7732 & 0.7682 & 0.7682 & 0.7680 & 0.1874 & 0.1320 & 0.1511 & 0.3011 & 22.2032 & 1.31439E-09 & 2595.74\\
ACS\_F814W & 0.8086 & 0.8059 & 0.7987 & 0.8142 & 0.8031 & 0.7990 & 0.8000 & 0.2891 & 0.1739 & 0.1856 & 0.2654 & 22.3391 & 1.15969E-09 & 2512.22\\
ACS\_F850LP & 0.9030 & 0.9016 & 0.8994 & 0.9060 & 0.9001 & 0.9006 & 0.9006 & 0.2534 & 0.1247 & 0.1210 & 0.1230 & 22.6773 & 8.49320E-10 & 2302.73\\
WFC3\_F218W & 0.2233 & 0.2229 & 0.2233 & 0.2242 & 0.2224 & 0.3635 & 0.3786 & 0.0570 & 0.0330 & 0.0340 & 0.0239 & 20.8017 & 4.77894E-09 & 791.70\\
WFC3\_F225W & 0.2379 & 0.2372 & 0.2374 & 0.2392 & 0.2365 & 0.2350 & 0.2335 & 0.0975 & 0.0467 & 0.0470 & 0.0404 & 20.9087 & 4.33036E-09 & 812.62\\
WFC3\_F336W & 0.3359 & 0.3355 & 0.3359 & 0.3366 & 0.3351 & 0.3371 & 0.3523 & 0.0714 & 0.0512 & 0.0550 & 0.1403 & 21.1944 & 3.32855E-09 & 1249.55\\
WFC3\_F390W & 0.3935 & 0.3924 & 0.4023 & 0.3956 & 0.3914 & 0.4522 & 0.3885 & 0.1247 & 0.0893 & 0.0948 & 0.1763 & 20.5639 & 5.94869E-09 & 3055.96\\
WFC3\_F438W & 0.4331 & 0.4326 & 0.4324 & 0.4340 & 0.4322 & 0.3884 & 0.3983 & 0.0849 & 0.0614 & 0.0673 & 0.1745 & 20.4103 & 6.85275E-09 & 4278.69\\
WFC3\_F475W & 0.4792 & 0.4774 & 0.4734 & 0.4829 & 0.4755 & 0.4763 & 0.4835 & 0.1671 & 0.1342 & 0.1481 & 0.2154 & 20.6805 & 5.34326E-09 & 4061.51\\
WFC3\_F555W & 0.5335 & 0.5308 & 0.5238 & 0.5389 & 0.5282 & 0.5309 & 0.5293 & 0.2846 & 0.1564 & 0.1579 & 0.1558 & 20.9843 & 4.03892E-09 & 3796.05\\
WFC3\_F606W & 0.5925 & 0.5887 & 0.5783 & 0.5999 & 0.5850 & 0.5917 & 0.5883 & 0.2561 & 0.2184 & 0.2298 & 0.2472 & 21.3170 & 2.97294E-09 & 3437.24\\
WFC3\_F625W & 0.6258 & 0.6241 & 0.6188 & 0.6291 & 0.6225 & 0.6222 & 0.6241 & 0.1762 & 0.1460 & 0.1573 & 0.2323 & 21.5086 & 2.49207E-09 & 3238.08\\
WFC3\_F775W & 0.7660 & 0.7648 & 0.7611 & 0.7684 & 0.7637 & 0.7642 & 0.7636 & 0.1752 & 0.1170 & 0.1455 & 0.1555 & 22.1828 & 1.33934E-09 & 2613.48\\
WFC3\_F814W & 0.8058 & 0.8030 & 0.7955 & 0.8117 & 0.8001 & 0.7956 & 0.7962 & 0.2746 & 0.1540 & 0.1518 & 0.1303 & 22.3259 & 1.17397E-09 & 2524.73\\
WFC3\_F098m & 0.9877 & 0.9864 & 0.9828 & 0.9903 & 0.9852 & 0.9872 & 0.9842 & 0.2085 & 0.1570 & 0.1694 & 0.3524 & 22.9160 & 6.81702E-10 & 2212.63\\
WFC3\_F105W & 1.0585 & 1.0551 & 1.0432 & 1.0652 & 1.0517 & 1.0560 & 1.0530 & 0.3290 & 0.2650 & 0.2917 & 0.4149 & 23.1459 & 5.51635E-10 & 2048.30\\
WFC3\_F125W & 1.2516 & 1.2486 & 1.2365 & 1.2576 & 1.2456 & 1.2455 & 1.2436 & 0.3427 & 0.2845 & 0.3005 & 0.4553 & 23.7673 & 3.11218E-10 & 1618.43\\
WFC3\_F140W & 1.3969 & 1.3922 & 1.3733 & 1.4061 & 1.3875 & 1.3804 & 1.3840 & 0.4438 & 0.3842 & 0.3941 & 0.4807 & 24.1788 & 2.13053E-10 & 1377.41\\
WFC3\_F160W & 1.5392 & 1.5370 & 1.5279 & 1.5436 & 1.5348 & 1.5322 & 1.5341 & 0.3302 & 0.2682 & 0.2874 & 0.4481 & 24.5692 & 1.48698E-10 & 1171.80\\
WFPC2\_F218W & 0.2214 & 0.2207 & 0.2205 & 0.2228 & 0.2200 & 0.3769 & 0.3781 & 0.0510 & 0.0451 & 0.0436 & 0.0026 & 20.7838 & 4.85820E-09 & 789.30\\
WFPC2\_F300W & 0.3013 & 0.2992 & 0.3039 & 0.3066 & 0.2972 & 0.2382 & 0.5499 & 0.1473 & 0.0857 & 0.0867 & 0.0116 & 21.0950 & 3.64742E-09 & 1089.44\\
WFPC2\_F450W & 0.4574 & 0.4556 & 0.4547 & 0.4608 & 0.4539 & 0.3830 & 0.3831 & 0.1730 & 0.0875 & 0.1078 & 0.0439 & 20.5907 & 5.80384E-09 & 4019.36\\
WFPC2\_F555W & 0.5468 & 0.5442 & 0.5373 & 0.5519 & 0.5417 & 0.5437 & 0.5416 & 0.2717 & 0.1456 & 0.1558 & 0.0604 & 21.0621 & 3.75986E-09 & 3714.54\\
WFPC2\_F606W & 0.6035 & 0.6001 & 0.5902 & 0.6101 & 0.5967 & 0.6009 & 0.5993 & 0.2677 & 0.1888 & 0.2002 & 0.1017 & 21.3763 & 2.81505E-09 & 3381.55\\
WFPC2\_F702W & 0.6945 & 0.6919 & 0.6841 & 0.6997 & 0.6893 & 0.6885 & 0.6884 & 0.2613 & 0.1666 & 0.1875 & 0.0920 & 21.8481 & 1.82293E-09 & 2910.78\\
WFPC2\_F814W & 0.8029 & 0.8002 & 0.7930 & 0.8087 & 0.7974 & 0.7930 & 0.7938 & 0.2859 & 0.1485 & 0.1455 & 0.0548 & 22.3159 & 1.18478E-09 & 2530.29\\
NIC2\_F110W & 1.1353 & 1.1235 & 1.0840 & 1.1575 & 1.1119 & 1.1232 & 1.1204 & 0.6564 & 0.4284 & 0.5272 & 0.1111 & 23.3646 & 4.50974E-10 & 1898.82\\
NIC2\_F160W & 1.6074 & 1.6030 & 1.5859 & 1.6159 & 1.5987 & 1.5932 & 1.5959 & 0.5042 & 0.3416 & 0.4013 & 0.1681 & 24.7184 & 1.29607E-10 & 1110.96\\
NIC3\_F110W & 1.1326 & 1.1200 & 1.0788 & 1.1561 & 1.1076 & 1.1142 & 1.1156 & 0.6568 & 0.4253 & 0.5883 & 0.1013 & 23.3547 & 4.55131E-10 & 1904.49\\
NIC3\_F160W & 1.6085 & 1.6042 & 1.5872 & 1.6169 & 1.5999 & 1.5965 & 1.5968 & 0.4967 & 0.3394 & 0.3987 & 0.1595 & 24.7213 & 1.29270E-10 & 1109.63\\
NIRCAM\_F070W & 0.7066 & 0.7046 & 0.6991 & 0.7114 & 0.7027 & 0.7053 & 0.7048 & 0.1896 & 0.1325 & 0.1600 & 0.1732 & 21.9113 & 1.71987E-09 & 2848.45\\
NIRCAM\_F090W & 0.9047 & 0.9025 & 0.8988 & 0.9093 & 0.9003 & 0.9039 & 0.9039 & 0.2383 & 0.1943 & 0.2101 & 0.2599 & 22.6730 & 8.52676E-10 & 2316.81\\
NIRCAM\_F115W & 1.1570 & 1.1543 & 1.1435 & 1.1624 & 1.1515 & 1.1526 & 1.1523 & 0.3155 & 0.2246 & 0.2683 & 0.2792 & 23.4729 & 4.08160E-10 & 1813.92\\
NIRCAM\_F140M & 1.4060 & 1.4053 & 1.4024 & 1.4074 & 1.4046 & 1.4064 & 1.4038 & 0.2186 & 0.1425 & 0.1478 & 0.2786 & 24.2262 & 2.03956E-10 & 1343.59\\
NIRCAM\_F150W & 1.5040 & 1.5007 & 1.4873 & 1.5104 & 1.4975 & 1.4974 & 1.4961 & 0.4093 & 0.3180 & 0.3371 & 0.3510 & 24.4711 & 1.62767E-10 & 1222.81\\
NIRCAM\_F150W2 & 1.7039 & 1.6588 & 1.4796 & 1.7864 & 1.6150 & 1.6383 & 1.5932 & 1.6908 & 1.1753 & 1.3255 & 0.3248 & 24.7096 & 1.30667E-10 & 1199.33\\
NIRCAM\_F162M & 1.6281 & 1.6272 & 1.6244 & 1.6297 & 1.6264 & 1.6255 & 1.6263 & 0.2522 & 0.1683 & 0.1714 & 0.2940 & 24.7932 & 1.20984E-10 & 1068.60\\
NIRCAM\_F164N & 1.6446 & 1.6445 & 1.6446 & 1.6446 & 1.6445 & 1.6386 & 1.6441 & 0.0602 & 0.0200 & 0.0179 & 0.1433 & 24.8561 & 1.14173E-10 & 1029.98\\
NIRCAM\_F182M & 1.8466 & 1.8452 & 1.8389 & 1.8494 & 1.8437 & 1.8423 & 1.8435 & 0.3335 & 0.2377 & 0.2460 & 0.3430 & 25.2723 & 7.78176E-11 & 883.75\\
NIRCAM\_F187N & 1.8739 & 1.8739 & 1.8737 & 1.8740 & 1.8739 & 1.8698 & 1.8731 & 0.0651 & 0.0237 & 0.0211 & 0.1650 & 25.3720 & 7.09931E-11 & 831.55\\
NIRCAM\_F200W & 1.9934 & 1.9886 & 1.9681 & 2.0028 & 1.9839 & 1.9803 & 1.9828 & 0.5611 & 0.4566 & 0.4717 & 0.4016 & 25.5530 & 6.00903E-11 & 792.68\\
NIRCAM\_F200W & 1.9934 & 1.9886 & 1.9681 & 2.0028 & 1.9839 & 1.9803 & 1.9828 & 0.5611 & 0.4566 & 0.4717 & 0.4016 & 25.5530 & 6.00903E-11 & 792.68\\
NIRCAM\_F210M & 2.0964 & 2.0955 & 2.0908 & 2.0982 & 2.0945 & 2.0956 & 2.0946 & 0.2955 & 0.2065 & 0.2090 & 0.3391 & 25.7709 & 4.91626E-11 & 720.06\\
NIRCAM\_F250M & 2.5038 & 2.5032 & 2.5006 & 2.5049 & 2.5027 & 2.5046 & 2.5026 & 0.2410 & 0.1800 & 0.1826 & 0.2979 & 26.4931 & 2.52780E-11 & 528.36\\
NIRCAM\_F277W & 2.7694 & 2.7618 & 2.7280 & 2.7845 & 2.7542 & 2.7476 & 2.7471 & 0.8977 & 0.6828 & 0.7111 & 0.3129 & 26.8793 & 1.77123E-11 & 450.64\\
NIRCAM\_F300M & 2.9908 & 2.9892 & 2.9819 & 2.9941 & 2.9875 & 2.9862 & 2.9852 & 0.5259 & 0.3153 & 0.3264 & 0.2353 & 27.2152 & 1.29998E-11 & 387.46\\
NIRCAM\_F322W2 & 3.2668 & 3.2320 & 3.0736 & 3.3335 & 3.1976 & 3.1789 & 3.1733 & 1.7937 & 1.3563 & 1.5827 & 0.3959 & 27.4646 & 1.03319E-11 & 359.99\\
NIRCAM\_F323N & 3.2370 & 3.2369 & 3.2368 & 3.2370 & 3.2369 & 3.2368 & 3.2375 & 0.0836 & 0.0385 & 0.0386 & 0.1466 & 27.5412 & 9.62799E-12 & 336.50\\
NIRCAM\_F335M & 3.3640 & 3.3621 & 3.3539 & 3.3676 & 3.3603 & 3.3592 & 3.3582 & 0.6118 & 0.3520 & 0.3608 & 0.2626 & 27.6991 & 8.32451E-12 & 313.89\\
NIRCAM\_F356W & 3.5768 & 3.5684 & 3.5287 & 3.5935 & 3.5600 & 3.5553 & 3.5532 & 1.0833 & 0.7811 & 0.8407 & 0.3766 & 27.9334 & 6.70861E-12 & 284.94\\
NIRCAM\_F405N & 4.0517 & 4.0517 & 4.0516 & 4.0517 & 4.0516 & 4.0475 & 4.0515 & 0.1108 & 0.0455 & 0.0460 & 0.1736 & 28.5039 & 3.96684E-12 & 217.21\\
NIRCAM\_F410M & 4.0844 & 4.0822 & 4.0723 & 4.0887 & 4.0801 & 4.0791 & 4.0790 & 0.7083 & 0.4379 & 0.4375 & 0.3096 & 28.5111 & 3.94060E-12 & 219.05\\
NIRCAM\_F430M & 4.2818 & 4.2813 & 4.2785 & 4.2829 & 4.2807 & 4.2806 & 4.2811 & 0.3664 & 0.2277 & 0.2312 & 0.3109 & 28.7127 & 3.27266E-12 & 200.09\\
NIRCAM\_F444W & 4.4157 & 4.4040 & 4.3496 & 4.4392 & 4.3923 & 4.4440 & 4.3830 & 2.6004 & 1.0316 & 1.1055 & 0.2042 & 28.8119 & 2.98700E-12 & 193.24\\
NIRCAM\_F460M & 4.6293 & 4.6285 & 4.6229 & 4.6308 & 4.6276 & 4.6253 & 4.6192 & 0.4007 & 0.2288 & 0.2323 & 0.2433 & 29.0433 & 2.41358E-12 & 172.47\\
NIRCAM\_F466N & 4.6545 & 4.6544 & 4.6540 & 4.6545 & 4.6544 & 4.6496 & 4.6544 & 0.1311 & 0.0536 & 0.0520 & 0.1347 & 29.0995 & 2.29198E-12 & 165.62\\
NIRCAM\_F470N & 4.7078 & 4.7078 & 4.7078 & 4.7079 & 4.7078 & 4.7054 & 4.7077 & 0.1196 & 0.0510 & 0.0495 & 0.1299 & 29.1127 & 2.26424E-12 & 167.39\\
NIRCAM\_F480M & 4.8181 & 4.8167 & 4.8094 & 4.8206 & 4.8154 & 4.8314 & 4.8197 & 0.5449 & 0.3073 & 0.3145 & 0.2253 & 29.2044 & 2.08078E-12 & 161.03\\
MIRI\_F560W & 5.6462 & 5.6362 & 5.5880 & 5.6661 & 5.6262 & 5.6138 & 5.6161 & 1.6516 & 0.9980 & 1.1178 & 0.1893 & 29.8562 & 1.14167E-12 & 120.97\\
MIRI\_F770W & 7.6669 & 7.6428 & 7.5260 & 7.7145 & 7.6188 & 7.5950 & 7.5977 & 2.4617 & 1.9647 & 2.1026 & 0.2962 & 31.1385 & 3.50433E-13 & 68.28\\
MIRI\_F1000W & 9.9694 & 9.9544 & 9.8806 & 9.9994 & 9.9394 & 9.9255 & 9.9272 & 2.4395 & 1.7910 & 1.8730 & 0.2867 & 32.2809 & 1.22357E-13 & 40.44\\
MIRI\_F1130W & 11.3111 & 11.3087 & 11.2962 & 11.3161 & 11.3062 & 11.3035 & 11.3050 & 1.5212 & 0.7336 & 0.7128 & 0.1577 & 32.8409 & 7.30515E-14 & 31.16\\
MIRI\_F1500W & 15.0929 & 15.0651 & 14.9272 & 15.1485 & 15.0373 & 15.0153 & 15.0107 & 4.3004 & 2.9217 & 3.1126 & 0.2387 & 34.0608 & 2.37513E-14 & 17.98\\
MIRI\_F1800W & 18.0088 & 17.9865 & 17.8760 & 18.0536 & 17.9641 & 17.9332 & 17.9422 & 4.7688 & 2.9569 & 2.9851 & 0.2070 & 34.8322 & 1.16713E-14 & 12.59\\
MIRI\_F2100W & 20.8425 & 20.7950 & 20.5607 & 20.9372 & 20.7476 & 20.7067 & 20.6991 & 7.1237 & 4.5749 & 4.6813 & 0.1953 & 35.4510 & 6.60100E-15 & 9.52\\
MIRI\_F2550W & 25.4081 & 25.3639 & 25.1519 & 25.4992 & 25.3197 & 25.2881 & 25.2803 & 8.0027 & 3.6615 & 3.4294 & 0.1019 & 36.3163 & 2.97500E-15 & 6.38\\
NIRISS\_F090W & 0.9058 & 0.9031 & 0.8985 & 0.9134 & 0.9004 & 0.9071 & 0.9095 & 0.2404 & 0.1833 & 0.2093 & 0.5295 & 22.6748 & 8.51330E-10 & 2316.26\\
NIRISS\_F115W & 1.1528 & 1.1499 & 1.1388 & 1.1588 & 1.1470 & 1.1490 & 1.1491 & 0.3180 & 0.2499 & 0.2699 & 0.5486 & 23.4581 & 4.13773E-10 & 1824.86\\
NIRISS\_F140M & 1.4054 & 1.4044 & 1.4010 & 1.4075 & 1.4035 & 1.4048 & 1.4034 & 0.2236 & 0.1424 & 0.1481 & 0.4086 & 24.2231 & 2.04526E-10 & 1345.65\\
NIRISS\_F150W & 1.4970 & 1.4936 & 1.4797 & 1.5040 & 1.4902 & 1.4907 & 1.4887 & 0.4145 & 0.3160 & 0.3399 & 0.4856 & 24.4520 & 1.65658E-10 & 1232.74\\
NIRISS\_F158M & 1.5851 & 1.5825 & 1.5704 & 1.5895 & 1.5799 & 1.5766 & 1.5739 & 0.8980 & 0.1990 & 0.2011 & 0.1222 & 24.6817 & 1.34061E-10 & 1119.85\\
NIRISS\_F200W & 1.9979 & 1.9930 & 1.9714 & 2.0077 & 1.9880 & 1.9857 & 1.9865 & 0.5681 & 0.4225 & 0.4741 & 0.5126 & 25.5611 & 5.96423E-11 & 790.19\\
NIRISS\_F277W & 2.7737 & 2.7641 & 2.7110 & 2.7910 & 2.7545 & 2.7242 & 2.7463 & 2.4443 & 0.6915 & 0.7281 & 0.2001 & 26.8736 & 1.78060E-11 & 453.79\\
NIRISS\_F356W & 3.6036 & 3.5926 & 3.5326 & 3.6243 & 3.5817 & 3.4346 & 3.5737 & 1.1841 & 0.9093 & 0.9242 & 0.5690 & 27.9544 & 6.58052E-12 & 283.31\\
NIRISS\_F380M & 3.8258 & 3.8229 & 3.7742 & 3.8284 & 3.8199 & 3.3856 & 3.8125 & 0.4586 & 0.2050 & 0.2056 & 0.3234 & 28.2278 & 5.11523E-12 & 249.36\\
NIRISS\_F430M & 4.2822 & 4.2792 & 4.2303 & 4.2848 & 4.2762 & 4.2904 & 4.2710 & 0.3414 & 0.2016 & 0.2135 & 0.3908 & 28.6952 & 3.32612E-12 & 203.16\\
NIRISS\_F444W & 4.4400 & 4.4270 & 4.3587 & 4.4653 & 4.4139 & 3.9705 & 4.4054 & 1.4206 & 1.0923 & 1.1403 & 0.5291 & 28.8293 & 2.93953E-12 & 192.16\\
NIRISS\_F480M & 4.8147 & 4.8113 & 4.7529 & 4.8181 & 4.8080 & 4.8294 & 4.7985 & 0.5192 & 0.2968 & 0.3026 & 0.3206 & 29.1856 & 2.11727E-12 & 163.49\\
OMEGACAM\_u & 0.3594 & 0.3590 & 0.3632 & 0.3602 & 0.3585 & 0.3242 & 0.3661 & 0.0851 & 0.0461 & 0.0527 & 0.1078 & 21.0647 & 3.75085E-09 & 1612.09\\
OMEGACAM\_g & 0.4751 & 0.4735 & 0.4702 & 0.4783 & 0.4719 & 0.3788 & 0.4771 & 0.1818 & 0.1150 & 0.1317 & 0.3397 & 20.6584 & 5.45315E-09 & 4077.99\\
OMEGACAM\_r & 0.6289 & 0.6276 & 0.6233 & 0.6316 & 0.6263 & 0.6260 & 0.6264 & 0.1963 & 0.1275 & 0.1351 & 0.3237 & 21.5293 & 2.44507E-09 & 3212.63\\
OMEGACAM\_i & 0.7508 & 0.7495 & 0.7453 & 0.7535 & 0.7482 & 0.7504 & 0.7491 & 0.2451 & 0.1143 & 0.1258 & 0.1920 & 22.1133 & 1.42790E-09 & 2675.72\\
OMEGACAM\_z & 0.8847 & 0.8842 & 0.8840 & 0.8856 & 0.8837 & 0.8822 & 0.8833 & 0.1677 & 0.0606 & 0.0530 & 0.0945 & 22.6336 & 8.84201E-10 & 2305.75\\
VIRCAM\_Z & 0.8950 & 0.8899 & 0.8815 & 0.9252 & 0.8849 & 0.8903 & 0.8975 & 1.4202 & 0.0929 & 0.0973 & 0.0477 & 22.6456 & 8.74504E-10 & 2310.18\\
VIRCAM\_Y & 1.0274 & 1.0253 & 1.0204 & 1.0363 & 1.0232 & 1.0250 & 1.0287 & 0.3265 & 0.0905 & 0.0924 & 0.1933 & 23.0396 & 6.08351E-10 & 2133.31\\
VIRCAM\_H & 1.2549 & 1.2535 & 1.2480 & 1.2586 & 1.2520 & 1.2502 & 1.2539 & 0.2619 & 0.1624 & 0.1720 & 0.4769 & 23.7910 & 3.04518E-10 & 1595.92\\
VIRCAM\_J & 1.6453 & 1.6430 & 1.6339 & 1.6499 & 1.6407 & 1.6374 & 1.6373 & 0.4530 & 0.2797 & 0.2905 & 0.5340 & 24.8212 & 1.17897E-10 & 1061.59\\
VIRCAM\_Ks & 2.1521 & 2.1494 & 2.1349 & 2.1567 & 2.1468 & 2.1840 & 2.1417 & 1.3215 & 0.2894 & 0.3078 & 0.1869 & 25.8668 & 4.50069E-11 & 693.60\\
SkyMapper\_u & 0.3616 & 0.3590 & 0.3685 & 0.3696 & 0.3565 & 0.3475 & 0.3267 & 0.4341 & 0.0456 & 0.0431 & 0.1040 & 21.1729 & 3.39505E-09 & 1459.58\\
SkyMapper\_v & 0.3837 & 0.3836 & 0.3874 & 0.3841 & 0.3834 & 0.3817 & 0.3831 & 0.0649 & 0.0318 & 0.0310 & 0.4905 & 20.6070 & 5.71726E-09 & 2805.75\\
SkyMapper\_g & 0.5099 & 0.5075 & 0.5016 & 0.5148 & 0.5051 & 0.5088 & 0.5044 & 0.2595 & 0.1477 & 0.1570 & 0.5693 & 20.8530 & 4.55807E-09 & 3916.05\\
SkyMapper\_r & 0.6157 & 0.6138 & 0.6078 & 0.6195 & 0.6120 & 0.6131 & 0.6134 & 0.2395 & 0.1524 & 0.1582 & 0.6359 & 21.4523 & 2.62480E-09 & 3298.97\\
SkyMapper\_i & 0.7778 & 0.7768 & 0.7734 & 0.7799 & 0.7758 & 0.7757 & 0.7750 & 0.1896 & 0.1202 & 0.1400 & 0.6336 & 22.2365 & 1.27465E-09 & 2565.56\\
SkyMapper\_z & 0.9159 & 0.9143 & 0.9119 & 0.9191 & 0.9128 & 0.9271 & 0.9206 & 0.2445 & 0.1110 & 0.0849 & 0.4540 & 22.7153 & 8.20138E-10 & 2286.99\\
\enddata
\end{deluxetable}
\end{longrotatetable}

\end{document}